
\def\scalehalf{
  \font\tenrm=cmr10 scaled \magstephalf
  \font\tenbf=cmbx10 scaled \magstephalf
  \font\tenit=cmti10 scaled \magstephalf
  \font\tensl=cmsl10 scaled \magstephalf
  \font\tentt=cmtt10 scaled \magstephalf
  \font\tenex=cmex10 scaled \magstephalf
  \font\teni=cmmi10 scaled \magstephalf
  \font\tensl=cmsl10 scaled \magstephalf
  \font\tensy=cmsy10 scaled \magstephalf
  \font\seveni=cmmi8
  \font\sevenrm=cmr8
  \font\sevensy=cmsy8
  \font\fivei=cmmi6
  \font\fiverm=cmr6
  \font\fivesy=cmsy6
  \font\rm=cmr10 scaled \magstephalf
  \font\bf=cmbx10 scaled \magstephalf
  \font\it=cmti10 scaled \magstephalf
  \font\sl=cmsl10 scaled \magstephalf
  \font\tt=cmtt10 scaled \magstephalf
  \normalbaselineskip 13truept
  \mathfamilydefs}
\def\mathfamilydefs{
  \def\rm{\fam0 \tenrm}
  \def\it{\fam\itfam \tenit}
  \def\sl{\fam\slfam \tensl}
  \def\bf{\fam\bffam \tenbf}
  \def\tt{\fam\ttfam \tentt}
  \def\mit{\fam1 }
  \def\cal{\fam2 }
  \textfont0=\tenrm  \scriptfont0=\sevenrm  \scriptscriptfont0=\fiverm
  \textfont1=\teni  \scriptfont1=\seveni  \scriptscriptfont1=\fivei
  \textfont2=\tensy  \scriptfont2=\sevensy  \scriptscriptfont2=\fivesy
  \textfont\itfam=\tenit
  \textfont\slfam=\tensl
  \textfont\bffam=\tenbf \scriptfont\bffam=\sevenbf
       \scriptscriptfont\bffam=\fivebf
  \textfont\ttfam=\tentt
  \normalbaselines\tenrm}
  
\font\eightrm=cmr8
\font\eighti=cmmi8
\skewchar\eighti='177
\font\eightsy=cmsy8
\skewchar\eightsy='60
\font\eightit=cmti8
\font\eightsl=cmsl8
\font\eightbf=cmbx8
\font\eighttt=cmtt8
\def\eightpoint{\textfont0=\eightrm \scriptfont0=\fiverm
                \def\rm{\fam0\eightrm}\relax
                \textfont1=\eighti \scriptfont1=\fivei
                \def\mit{\fam1}\def\oldstyle{\fam1\eighti}\relax
                \textfont2=\eightsy \scriptfont2=\fivesy
                \def\cal{\fam2}\relax
                \textfont3=\tenex \scriptfont3=\tenex
                \def\it{\fam\itfam\eightit}\relax
                \textfont\itfam=\eightit
                \def\sl{\fam\slfam\eightsl}\relax
                \textfont\slfam=\eightsl
                \def\bf{\fam\bffam\eightbf}\relax
                \textfont\bffam=\eightbf \scriptfont\bffam=\fivebf
                \def\tt{\fam\ttfam\eighttt}\relax
                \textfont\ttfam=\eighttt
                \setbox\strutbox=\hbox{\vrule
                     height7pt depth2pt width0pt}\baselineskip=10pt
                \rm}
\font \bigone=cmbx10 scaled\magstep1

\def\d3k{{\displaystyle d{\bf k} \over \displaystyle (2\pi)^3}}

\def\hang{\par\hangindent=\parindent\@}
\def\@{\noindent}

%
%
\newcount\eqnumber
\eqnumber=1
\def\step#1{\global\advance#1 by 1}
\def\neweq{{\rm\the\eqnumber}\step{\eqnumber}}
\def\eqnew{\eqno(\neweq)}
{\scalehalf
\newcount\startpage
\startpage=1
\pageno=\startpage
\baselineskip 16pt
\voffset=-0.5truecm
\overfullrule=0pt
\tolerance=1500
\centerline{\bf{\bigone CONSTRAINING PEAKS IN GAUSSIAN PRIMORDIAL
DENSITY FIELDS:}}
\medskip
\centerline{\bf{\bigone AN APPLICATION OF THE HOFFMAN-RIBAK METHOD}}
\bigskip
\centerline{\bf Rien van de Weygaert$^{1,2,3}$ \& Edmund Bertschinger$^4$}
\medskip
\vbox{\noindent\eightpoint
\centerline{$^{1}$ Kapteyn Instituut, University of Groningen,
P.O. Box 800, 9700 AV Groningen, the Netherlands}
\centerline{$^{2}$ Max-Planck-Institut f\"ur Astrophysik,
Karl-Schwarzschild-Stra{\ss}e 1, Garching bei M\"unchen, Germany}
\centerline{$^{3}$ Canadian Institute for Theoretical Astrophysics,
60 St. George Street, Toronto, Ontario M5S 1A7, Canada}
\centerline{$^{4}$  Department of Physics, Massachusetts Institute of
Technology, Cambridge, MA 02139, U.S.A.}
\smallskip
\centerline{Email: weygaert@astro.rug.nl, bertschinger@mit.edu}
}
\bigskip
\centerline{\bf ABSTRACT}
\medskip
\vbox{\baselineskip 12pt
We develop an algorithm for setting up initial Gaussian random density and
velocity
fields containing one or more peaks or dips, in an arbitrary cosmological
scenario. The
intention is to generate appropriate initial conditions for cosmological N-body
simulations
that focus on the evolution of the progenitors of the present-day galaxies and
clusters. The
procedure is an application of the direct and accurate prescription of Hoffman
\& Ribak (1991) for
generating constrained random fields.

For each peak a total of 21 physical characteristics can be specified,
including its scale,
position, density Hessian, velocity, and velocity gradient.  The velocity (or,
equivalently, gravity) field constrants are based on a generalization of the
formalism
developed by Bardeen et al. (1986).  The resulting density field is sculpted
such that it
induces the desired amount of net gravitational and tidal forces.

We provide a detailed mathematical presentation of the formalism.  Afterwards
we provide
analytical estimates of the likelihood of the imposed
constraints. Amongst others, it is shown that the tidal field has a strong
tendency to align itself along the principal axes of the mass tensor. The
method is illustrated
by means of some concrete examples. In addition to the illustration of
constraint-field
correlation functions and how they add up to the mean fields, followed by
illustrations of the
variance characteristics of field realizations, we concentrate in particular on
the
consequences of imposing gravitational field constraints (or, equivalent in the
linear regime for growing mode fluctuations, peculiar velocity field
constraints).}

\medskip
\@\medskip
\@{\it Subject headings:} Cosmology : theory -- Galaxies: clustering --
large-scale structure of the Universe -- Methods: numerical

\medskip
\@{\bf {\bigone 1. Introduction}}
\smallskip
\@In the standard scenario of structure formation galaxies and the
large-scale structure form through the growth of primordial
density perturbations. These perturbations take the form of a homogeneous
and isotropic random process. In most cases these cosmological density
fields are assumed to be Gaussian random fields.

In these density fields the regions around local maxima and minima are of
particular interest during the evolution of the perturbation field. The
first collapsed structures form generally near (but are not coincident with,
Bertschinger \& Jain 1994)
density peaks, making density maxima the progenitors of objects like
galaxies and clusters. On the other hand, the
minima will be the centres of expanding voids. The properties of peaks in
Gaussian random fields have been described extensively in the literature.
In order to identify an object of a certain size and mass in a Gaussian random
field one discards smaller scale objects from consideration. This is achieved
by filtering the density field on an appropriate scale to reflect the linear
evolution of the proto-objects. While in some scenarios the filter function is
a consequence of a simple phenomenon (e.g. free-streaming of neutrinos in a
Hot Dark Matter scenario) in other cases one is forced to invoke an artificial
filter to approximate the complicated processes of hierarchical merging
(e.g. in the Cold Dark Matter scenario).

A description of the properties of these filtered fields was given by
Doroshkevich (1970), Peacock and Heavens (1985), and Bardeen et al. (1986;
hereafter BBKS). Beside global parameters such as the number density and
spatial correlations of peaks found in these filtered fields they also
derived the distribution of their height, shape and orientation. Furthermore,
BBKS derived the mean and variance of the density profiles around peaks.
As soon as these structures enter the nonlinear regime the coupling of
modes breaks down the above approach of filtering. To investigate the further
evolution one is therefore forced to resort to N-body simulations. However, in
order to follow the evolution of a particular object one needs to be able to
start off with a primordial density field containing such an object.

Unfortunately, the methods of BBKS apply only to point processes and
cannot be used to construct an actual sample of a density profile around
a peak with predetermined parameters such as peak height, shape and
orientation. The usual approach is therefore to generate an unconstrained
realization of a Gaussian field and then to search for peaks or regions that
satisfy the desired constraints. In many instances this is an inefficient
approach. For example, giant clusters or voids will be so rare that
either many samples have to be generated or that a large box needs to be used
to ensure that the object is indeed present in the simulation volume. The
latter will yield a severely degraded resolution which conflicts with the
desire to describe these objects in as much detail as possible. Similar
considerations apply when many properties need to be specified to obtain
the desired object, even while the corresponding additional constraints do not
represent unlikely values. By being able to specify beforehand some of the
properties and to ensure the presence of such a peak or region in the
simulation volume the required effort will be minimized. Simultaneously, the
resolution will be maximized. Potentially the most important advantage of this
approach is that the influence of several physical quantities on the evolution
of structures can be studied systematically by generating realizations
wherein one or more constraints have various values.

The fundamental theory of these constrained random fields was set forth by
Bertschinger (1987). He generalized the treatment used by BBKS
to give a full statistical description of a Gaussian random field subjected to
constraints. Based on these principles he presented a method to correctly
sample the probability distribution of the density field subject to linear
constraints. This method, however, is rather elaborate and inefficient in
its implementation, involving a simulated annealing technique. Although it
is useful for generating initial conditions subject to a few constraints
(see e.g. Van de Weygaert \& Van Kampen 1993), it quickly becomes
prohibitively slow for more than two constraints.
Looking for a more efficient procedure, Binney and Quinn (1991) showed that
Bert\-schinger's problem simplifies considerably when the random field
is expanded in spherical harmonics rather than in a plane wave basis.
In the case of a localised set of constraints, such as the presence and shape
of a peak at the centre of the box, the problem can then be solved exactly
instead of iteratively. However, their algorithm is essentially restricted to
the case of quite localised constraints defined around an obvious centre of
symmetry.

The breakthrough in the construction of constrained random fields came with
the publication by Hoffman \& Ribak (1991, hereafter HR). They realised that
for any constraint that is a linear functional of the field the problem can be
solved exactly in an elegant and simple manner, without having to invoke
complicated iterative techniques. Their method makes it possible to generate
initial conditions for N-body simulations that obey a few hundred
constraints, e.g. those imposed by the observable universe
(see Ganon \& Hoffman 1993).

This paper contains a description of the fundamentals and implementation of
a specific cosmological application of the method proposed by
Hoffman \& Ribak (1991). This application consists of the
generation of an initial density and velocity field containing one or more
density peaks in a simulation box. Apart from being able to determine the
location and the scale of the peak, we can specify the central density of
the peak, as well as the compactness, shape and orientation of the density
field in the immediate surroundings of the peak. In addition, the total
matter distribution can be sculpted such that it subjects the peak to a
desired amount of net gravitational and tidal forces.
In practice, the computer algorithm generates samples of these constrained
Gaussian random fields on a lattice, using Monte Carlo techniques.
Nearly all relevant calculations are done in Fourier transform space.
Some results of cosmological studies based on these constrained initial
conditions are presented by Van Haarlem \& Van de Weygaert (1993),
Van de Weygaert \& Babul (1994, 1995).

In this paper, we start with some basic concepts of Gaussian random
fields followed by a treatment of the fundamental theory of constrained
Gaussian random fields in section~2. The Hoffman-Ribak method for the
construction of constrained random fields is described in section~3,
followed by a description of our Fourier space implementation. In section~4,
we present our application of this method to the generation of peaks, deriving
constraint kernels for the various peak quantities. In addition, we provide
prescriptions for the probability of the imposed constraints. A
realization of a random density field with a constrained peak is presented
in section~5. Specifically, we will focus on the influence of imposing a
peculiar acceleration and a tidal field.
In section~6,  we will conclude with a summary and short discussion.

\medskip
\@{\bf {\bigone 2. Fundamentals of constrained Gaussian random fields}}
\smallskip
\@Although the paper by Hoffman and Ribak presents the essentials of
the simple direct method to construct samples of constrained random fields, it
does not provide its mathematical background. This can be obtained
extending our earlier treatments (Bertschinger 1987, Van de Weygaert 1991).
Therefore we will first summarize the necessary mathematical background in the
notation employed by HR before we get to the presentation of their method.

\medskip
\@{\bf{2.1 Gaussian random fields: basics}}
\smallskip
\@Consider a homogeneous and isotropic random field $f({\bf x})$ with zero
mean. The random field is defined by the set of $N$-point joint
probabilities,

$${\cal P}_N=P[f({\bf x}_1),f({\bf x}_2),\ldots,f({\bf x}_N)]\,df({\bf x}_1)
df({\bf x}_2) \cdots df({\bf x}_N),\eqnew$$

\@that the field $f$ has values in the range $f({\bf x}_j)$ to $f({\bf x}_j)+
df({\bf x}_j)$ for each of the $j=1,\ldots,N$, with $N$ an arbitrary integer
and specified positions ${\bf x}_1,{\bf x}_2,\ldots,{\bf x}_N$.

Here we restrict ourselves to the study of Gaussian random fields, whose
statistical properties are completely characterised by some power spectrum
(spectral density) or its Fourier transform, the
autocorrelation function.
There are both physical and statistical arguments in favour of the
assumption that the primordial density field in the Universe was indeed
of this nature. If the very early Universe went through an inflationary phase,
quantum fluctuations would generate small-amplitude curvature fluctuations.
The resulting density perturbation field is generally a Gaussian random
process with a nearly Harrison-Zel'dovich scale-invariant primordial power
spectrum. But even while inflation did not occur, the density field
$f({\bf x})$ will be nearly Gaussian in the rather general case that its
Fourier components ${\hat f}({\bf k})$ are independent and have random
phases (cf. Scherrer 1992).
The Fourier decomposition of the field at a specific location
${\bf x}$ can then be seen as the superposition of a large number of
independent random variables that are drawn from the same distribution.
By virtue of the central limit theorem the distribution of this field
will approach normality, and (at least) for small $N$ the multivariate
distribution ${\cal P}_N$ is multivariate normal (Gaussian):

$${\cal P}_N={\displaystyle \exp\left[-{\displaystyle {1 \over 2}}\,
\sum\nolimits_{i=1}^N\,\sum\nolimits_{j=1}^N\,f_i\,({\bf M} ^{-1})_{ij}
\,f_j\right] \over \displaystyle [(2\pi)^N\,(\det {\bf M})]^{1/2}}\,
\prod_{i=1}^N\,df_i\,,\eqnew$$

\@where ${\bf M}^{-1}$ is the inverse of the $N \times N$ covariance matrix
{\bf M}, the generalisation of the variance $\sigma^2$ in a
one-dimensional normal distribution. ${\bf M}$ is completely determined by
the autocorrelation function $\xi(r)$ if the field is a Gaussian random
field,

$$M_{ij} \equiv \langle f({\bf x}_i) f({\bf x}_j) \rangle =
\xi({\bf x}_i-{\bf x}_j)=\xi(|{\bf x}_i-{\bf x}_j|),\eqnew$$

\@Throughout this paper the brackets $\langle \dots \rangle$ denote an
ensemble average. The last relation in equation~(3) reflects the fact that our
field is a homogeneous and isotropic random process.
Since we can consider $f$ as an $N$-dimensional column vector, we can also
write the covariance matrix ${\bf M}$ in the convenient form

$${\bf M}=\langle f\,f^t\rangle,\eqnew$$

\@with $f^t$ the transpose of $f$. By taking the limit as $N \rightarrow
\infty$
with uniform spatial sampling,
the summations appearing in equation~(2) may be turned into
integrals. The result

$${\cal P}[f]=e^{-S[f]}\ {\cal D}\left[f\right]\,,\eqnew$$

\@is similar to the quantum-mechanical partition function in path integral
form, where $S$ is the action functional. Although there is no direct
connection with quantum field theory, $S$ will be referred to as the action.
The expression for the action $S$ for a Gaussian random field can be
obtained from equation~(2),

$$S[f]={\displaystyle {1 \over 2}}\,\int d{\bf x}_1\int d{\bf x}_2\,
f({\bf x}_1)K({\bf x}_1-{\bf x}_2)f({\bf x}_2)\,,\eqnew$$

\@where $K$ is the functional inverse of the correlation function $\xi$,

$$\int\,d{\bf x}\,K({\bf x}_1-{\bf x})\xi({\bf x}-{\bf x}_2)=
\delta_D({\bf x}_1-{\bf x}_2),\eqnew$$

\@and $\delta_D$ the Dirac delta function. The measure
${\cal D}\left[f\right]$ is most easily evaluated on a lattice,
where it is just the product of differentials $df_i$ divided by a
normalization constant.

Note that we use the notation ${\cal P}[f]$ to refer to an infinitesimal
probability with measure ${\cal D}\left[f\right]$; the probability density
is $\exp\left(-S[f]\right)$.  The square brackets in ${\cal P}[f]$
and $S[f]$ indicate that these are functionals, i.e., they map the
complete function $f({\bf x})$ to one number.

To compute expectation values $\langle A \rangle$ of properties (functionals)
of the random field $f({\bf x})$, such as the galaxy mass distribution or
the distribution of cluster shapes, one integrates the functional over all
possible density fields $f({\bf x})$, weighting each by the probability
from equation~(5),

$$\langle A \rangle = {\displaystyle \int A[f]\,e^{-S[f]}\ {\cal D}[f]
\over \displaystyle \int e^{-S[f]}\ {\cal D}[f]}.\eqnew$$

\@This is exactly analogous to the sum over histories or
paths in the Feynman path integral formulation of quantum mechanics (Feynman
and Hibbs 1965). As in quantum field theory, there are two practical
ways to evaluate cosmological path integrals: perturbation series and
Monte Carlo integration.

The perturbation series approach to path integrals, based on Feynmann diagrams,
is limited to a small number of applications, as it runs into difficulties
when cosmological structures become nonlinear. A more general way to evaluate
path integrals, which is adopted here, is by Monte Carlo integration. By
generating realisations $f_i$ of the density field, and evaluating the
corresponding values $A[f_i]$ of the quantity $A$, the mean of these values
is determined:

$$\langle A \rangle ={\displaystyle \sum\nolimits_i\,A[f_i] \over
\displaystyle \sum\nolimits_i\,1}.\eqnew$$

\@The subsequent non-linear evolution is treated by performing $N$-body
simulations of a specific realisation $f$. The central issue in this method
is the need to draw samples $f_i$ which have a probability distribution
proportional to $\exp(-S[f])$ (eq.~5).

\medskip
\@{\bf{2.2 Gaussian Random Fields: constraints}}
\smallskip
\@The complicating factor in generating Gaussian random density fields
subject to one or more constraints is that correlations couple all points
of the field with all other points. Therefore, instead of describing the
field in terms of an infinite product of one-dimensional probabilities,
one is forced to formulate the problem using infinite-dimensional
probability spaces (see eq.~5).

The strategy followed by Bertschinger (1987) is to incorporate the set
of constraints imposed on the density field $f({\bf x})$ in the action $S[f]$,
according to the definition in equation~(5). A realization of the constrained
density field is then obtained by properly sampling the resulting
distribution
function $\exp(-S[f])$. To make clear how the constraints are incorporated
in the action, we consider a field $f({\bf x})$ that is subject to a set of
$M$ constraints,

$$\Gamma=\{C_i\equiv C_i[f;{\bf x_i}]=c_i\,; \quad i=1,\dots,M\}.
\eqnew$$

\@The constraints are therefore imposed by forcing the field
$C_i[f;{\bf x}]$, $(i=1,\dots,M)$, a functional of the field $f({\bf x})$
as well as a function of the point ${\bf x}$, to have
the specific value $c_i$ at the position ${\bf x}_i$. The constraints $C_i$
are assumed to be linear functionals.
Examples of such functionals are the value of the field itself at the point
${\bf x}_{\alpha}$, the derivative of the field $f({\bf x})$ at the point
${\bf x}_{\beta}$, or a convolution over $f({\bf x})$ with some function
$g({\bf x})$,

$$\eqalign{C_{\alpha}[f;{\bf x}_{\alpha}]&=f({\bf x}_{\alpha})=c_{\alpha},\cr
C_{\beta}[f;{\bf x}_{\beta}]&={\partial \over \partial x} f({\bf x})
|_{{\bf x}_{\beta}} = c_{\beta},\cr
C_{\gamma}[f;{\bf x}_{\gamma}]&=\int\,g({\bf x}_\gamma-{\bf x})\,
f({\bf x})\,d{\bf x} = c_{\gamma}.}\eqnew$$

\@The constraints $C_{\alpha}$ and $C_{\beta}$ can be considered as
particular cases of a convolution of $f({\bf x})$ with functions
$g_{\alpha}$ and $g_{\beta}$ respectively,

$$\eqalign{g_{\alpha}({\bf x}_\alpha-{\bf x})&=
\delta_D({\bf x}_\alpha-{\bf x})\cr
g_{\beta}({\bf x}_\beta-{\bf x})&= {\partial \over \partial x}
\delta_D({\bf x}_\beta-{\bf x}).}\eqnew$$

\@A broad class of constraints can be considered as such, so that a
treatment of the constraints in the form of a convolution is not a serious
restriction. In particular, we will see in section~4 that the expressions for
the 10 constraints needed to specify the height, shape and orientation of a
peak in the filtered density field $f_F({\bf x})$, the 3 constraints to specify
its peculiar acceleration and the 5 constraints to specify its tidal field
can all be written as convolutions over the field $f({\bf x})$, with the
convolution functions $g$ depending on the kind of constraint.

\vskip 0.5truecm
Since we limit our fields $f({\bf x})$ to those that obey the set of $M$
constraints
$\Gamma$, the probability of possible realisations $f({\bf x})$ is the
conditional probability ${\cal P}[f({\bf x})|\Gamma]$,

$${\cal P}\left[f|\Gamma\right]={ {\cal P}\left[f,\Gamma
\right] \over {\cal P}\left[\Gamma\right]}={{\cal P}\left[f
\right] \over {\cal P}\left[\Gamma\right]}.\eqnew$$

\@The second equality follows because the constraints are linear functionals of
$f$, so that the joint probability space for $f$ and $\Gamma$ is the same as
the
probability space for $f$.
Because the constraints $C_i$ are linear functionals
the central limit theorem assures them to have a Gaussian probability
distribution when applied on a Gaussian field $f({\bf x})$. The covariance
matrix ${\bf Q}$ of the constraints' probability distribution can be
expressed as (cf. eq.~4),

$${\bf Q}= \langle C\,C^t \rangle,\eqnew$$

\@where $C$ is the $M$-dimensional column vector with elements $C_i$, and $C^t$
its transpose. The joint probability ${\cal P}\left[\Gamma\right]$ for the
set of
constraints $\Gamma$ is therefore the following multivariate
Gaussian distribution (cf. eq.~2),

$${\cal P}\left[\Gamma\right]={\displaystyle \exp\left[-{\displaystyle
{1 \over  2}}\,
\sum\nolimits_{i=1}^{M}\,\sum\nolimits_{j=1}^{M}\,C_i\,
({\bf Q}^{-1})_{ij}\,C_j\right] \over \displaystyle [(2\pi)^{M}\,
(\det {\bf Q})]^{1/2}}\,\prod_{i=1}^{M}\,dC_i\,,\eqnew$$

\@or, in a more concise notation,

$${\cal P}\left[\Gamma\right]=\exp\left(-{\displaystyle {1 \over 2}} C^t
{\bf Q}^{-1} C\right) \,\,{\cal D}[\Gamma],\eqnew$$

\@where the measure ${\cal D}[\Gamma]$ is defined as

$${\cal D}[\Gamma]={\displaystyle 1 \over \displaystyle
[(2\pi)^{M}\,(\det {\bf Q})]^{1/2}}\,\prod_{i=1}^{M}\,dC_i.\eqnew$$

\@When each field $f({\bf x})$ is represented by its value at $N$ points
(e.g. in a discrete computer representation) we can picture the problem in a
geometrical way. The fields $f({\bf x})$ can be considered as $N$-dimensional
vectors
$(f_1,\dots,f_N)$. The constraint set $\Gamma$ carves out an
$(N-M)$-dimensional hypersurface in this $N$-dimensional vector space,
consisting of all fields obeying these constraints. In other words, the set
$\Gamma$ is an $(N-M)$-dimensional hypersurface, in particular a hyperplane
when the constraints are linear.

The expression for the conditional probability of the field $f({\bf x})$
given the set of constraints $\Gamma$, ${\cal P}[f|\Gamma]$, follows after
inserting equations~(5), (6) and (16) into equation~(13),

$${\cal P}\left[f|\Gamma\right]=\exp\left[-{\displaystyle {1 \over 2}}
\left(\,\int \int\,f({\bf x}_1)K({\bf x}_1-{\bf x}_2)f({\bf x}_2)
\,\,d{\bf x}_1\,d{\bf x}_2\, - \,C^t
{\bf Q}^{-1} C\right)\right]\,\,{\displaystyle {\cal D}[f] \over
{\cal D}[\Gamma]}.\eqnew$$

\@This result shows that the constraints $\Gamma$ are incorporated
into the formalism by a change of the action $S[f]$ to

$$2S[f]=\int \int\,f({\bf x}_1)K({\bf x}_1-{\bf x}_2)f({\bf x}_2)\,\,
d{\bf x}_1\,d{\bf x}_2\, - \,C^t {\bf Q}^{-1} C\eqnew$$

\@In Appendix A it is shown that this constrained action may be written in
a simple and revealing form,

$$2S[F]=\int d{\bf x}_1\int d{\bf x}_2\,F({\bf x}_1) K({\bf x}_1-
{\bf x}_2)F({\bf x}_2),\eqnew$$

\@where the residual field $F({\bf x})$ is defined as the difference between
a Gaussian field $f({\bf x})$ satisfying the constraint set $\Gamma$ and the
ensemble mean ${\bar f}({\bf x})$ of all these fields,

$$F({\bf x}) \equiv f({\bf x})-{\bar f}({\bf x}).\eqnew$$

\topinsert
\vbox{\noindent\eightpoint
{Figure 1.}
Illustration of the construction of a
constrained random field. The field contains two peaks, an elongated
one defined on a Gaussian scale of $4h^{-1}$ Mpc
at $[x,y]=[65.0,65.0]$ $h^{-1}$ Mpc, and a more compact one defined on a
Gaussian scale of $2h^{-1}$ Mpc at a position of $[x,y]=[35.0,35.0]$ $h^{-1}$
Mpc.
The corresponding mean constrained field $({\bar f})$ is shown in the top left
frame, to which the residual field $F=f-{\bar f}$ in the top right frame
is added to obtain the constrained random field realization $(f)$ shown
in the bottom frames. The left frame shows the field after smoothing with
a Gaussian filter with $R_f=2h^{-1}\,\hbox{Mpc}$, while the right frame is
the field after smoothing on a scale of $R_f=4h^{-1}\,\hbox{Mpc}$. The
fluctuation field has a standard cold dark matter spectrum ($\Omega=1.0,
h=0.5$).
}\endinsert

\@The conditional probability function can therefore be described as a
shifted
Gaussian around the ensemble mean field, ${\bar f}({\bf x})$ (see
Appendix A),

$${\bar f}({\bf x})=\langle f({\bf x})|\Gamma \rangle =
\xi_i({\bf x})\,\xi_{ij}^{-1}\,c_j,\eqnew$$

\@where summation over repeated indices is used. Thus, ${\bar f}({\bf x})$
is the ``most likely'' field satisfying the constraints and it equals the
``average density profile'' obtained by BBKS. More precisely, $f={\bar f}$ is
a stationary point of the action:

$${\displaystyle \delta S \over \displaystyle \delta f}=0\qquad
\hbox{for}\quad f = {\bar f}.\eqnew$$

\@In equation~(22) $\xi_i({\bf x})$ is the cross-correlation between the
field and the $i$th constraint $C_i[f;{\bf x}_i]$ while $\xi_{ij}$ is the
$(ij)^{th}$ element of the constraints' correlation matrix ${\bf Q}$,

$$\eqalign{\xi_i({\bf x})&=\langle f({\bf x})\, C_i\rangle,\cr
\xi_{ij}&=\langle C_i\,C_j\rangle.}\eqnew$$

\@If the constraints $C_i$ involve only the field itself at single points, like
$C_{\alpha}$ in equation~(11), both the correlation matrix $\xi_{ij}$ and
$\xi_i({\bf x})$ reduce to the two-point correlation function $\xi({\bf x})$,

$$\eqalign{\xi_i({\bf x})&=\langle f({\bf x})\,f({\bf x}_i) \rangle =
\xi(|{\bf x}_i-{\bf x}|),\cr
\xi_{ij}&=\langle f({\bf x}_i)\,f({\bf x}_j) \rangle = \xi(|{\bf x}_i
- {\bf x}_j|). }\eqnew$$

\@{\it In effect, the residual field $F({\bf x})$ provides random noise which
is added to the signal ${\bar f}({\bf x})$, which is completely fixed by the
imposed set of constraints $\Gamma$}. Generating a sample $f({\bf x})$
obeying the constraints $\{C_i[f;{\bf x}_i]=c_i;\ \ i=1,\dots,M\}$
therefore consists
of constructing ${\bar f}$ from $C_i$ and $c_i$ according to equation~(22),
subsequently generating the noise $F({\bf x})$, and adding them:

$$f({\bf x})={\bar f}({\bf x})\,+\,F({\bf x})\,=\,\xi_i({\bf x})
\xi_{ij}^{-1} c_i\,+\,F({\bf x}).\eqnew$$

\@Notice that the residual field $F$ is a Gaussian field because it is the
difference between two Gaussian fields. The whole problem of constructing
a constrained random field has now been reduced to a proper sampling of
$F$. This is complicated by the fact that $F({\bf x})$ is not
entirely random but subject to the set of $M$ constraints $\Gamma_0$:

$$\Gamma_0 \equiv \{C_i[f;{\bf x}_i]=0\,;\quad i=1,\dots,M\}.\eqnew$$

\@This follows directly from the fact that the constraints $C_i$ are linear
functionals and $F$ is the difference between two fields,

$$C_i\left[F\right]=C_i\left[f-{\bar f}\right]=
C_i\left[f\right]-C_i\left[{\bar f}\right]=c_i-c_i=0.\eqnew$$

\@This fact is independent of the numerical values $\{c_i\}$ of the
constraints $\Gamma$ imposed on the field $f({\bf x})$.

\topinsert
\vbox{\noindent\eightpoint
{Figure 2.}
Linear density profiles along the central $x$-axis of the field shown in figure
1.
On the left, the field has been smoothed using a Gaussian filter with
$R_f=2h^{-1}\,
\hbox{Mpc}$. The solid line shows the constrained field $(f)$. The dotted line
is the mean
field $({\bar f})$ and the dashed line the residual field $F=f-{\bar f}$. On
the right the
same field, but now after filtering on a scale of $4h^{-1}\,\hbox{Mpc}$.

An illustration of the sketched constrained field construction procedure, based
on
equation~(26), is provided by figure~1. Note that both the original
Bertschinger
prescription (1987) and the Hoffman-Ribak procedure (1991) are based on this
equation
(the particular realization in figure~1 has been generated with the
Hoffman-Ribak code
described in this paper). The fluctuation field in the $100h^{-1}\,\hbox{Mpc}$
box has
a standard cold dark matter spectrum $(\Omega=1.0,h=0.5)$ and contains two
peaks of different
shape and
scale, a spherical $4 \sigma_0(2h^{-1}\,\hbox{Mpc})$ overdensity and an
elongated
$3 \sigma_0(4h^{-1}\,\hbox{Mpc})$ overdensity. Density contour maps (filtered
on a
scale of $2h^{-1}\,\hbox{Mpc}$) of the mean field ${\bar f}$ defined by this
constraint
(top left), an accompanying residual field realisation $F$ (top right) and the
resulting constrained field $f$ (bottom left) are shown in slices of width
$1/20th$ of
the boxsize taken along the $z$-direction. The slices pass through the centre
of the box.
Figure~1d shows the constrained field $f$ smoothed on a scale of $4\,h^{-1}\,
\hbox{Mpc}$. A good idea of the relative amplitudes of the mean, residual and
constrained field in figure~1 can be obtained from linear density profiles
through the
density field. Figure~2 shows such profiles, taken along the central $x$-axis,
passing
through the outskirts of both peaks. The left figure corresponds to the density
field at a Gaussian filtering scale of $2h^{-1}\,\hbox{Mpc}$, while the right
figure
has a Gaussian smoothing scale of $4h^{-1}\,\hbox{Mpc}$. The dotted line is the
mean
field ${\bar f}$, the dashed line the residual field $F$, and the solid line
the
superposition of the two, the constrained field realization $f$.
}\endinsert

\medskip
\@{\bf {\bigone 3. Sampling constrained Gaussian random fields}}
\smallskip
\@Application of the construction procedure based on equation~(26)
requires the ability to properly sample $\exp(-S[F])$ for the random field
$F({\bf x})$. The sampling procedure forms the core of any constrained
random field algorithm, and determines its effectiveness and reliability.
The sampling is carried out most conveniently in Fourier space,
where the action $S[F]$ is diagonalized (appendix~B),

$$S[F]=\int \d3k\,
{\displaystyle |{\hat F}({\bf k})|^2 \over \displaystyle 2 P(k)}
\,,\eqnew$$

\@where ${\hat F}({\bf k})$ is the Fourier transform of the residual field
$F({\bf x})$,

$$F({\bf x})=\int \d3k
\,{\hat F}({\bf k})\,e^{-i{\bf k}\cdot{\bf x}}\,,\eqnew$$

\@and $P(k)$ the power spectrum of the field (see eq.~41 for the formal
definition).
Note that in this paper we adopt a different Fourier transform than
Bertschinger (1987,
1992).

In the case of an unconstrained field, for which
$F({\bf x})=f({\bf x})$, all harmonics ${\hat F}({\bf k})$ are mutually
independent and normally distributed. This makes sampling relatively easy.
However, for a constrained field the residual field is subject to the
constraints $\Gamma_0$ (eq.~27), so that its Fourier components are no
longer mutually independent. The coupling of the different Fourier
modes turns the sampling of the action $S[F]$ into a non-trivial problem.
In earlier work we (Bertschinger 1987, Van de Weygaert 1991) accomplished
the sampling
of the residual field, carried out in discrete Fourier space
${\hat F}({\bf k}_j)$, by means of an iterative ``simulated annealing''
technique. The action was sampled by means of a Markov chain, starting with
an initial guess for the harmonics and updating them iteratively, each update
depending only on values of the most recent estimate. After a number of
iterations the Markov chain relaxes to a steady state with ${\hat F}$
correctly sampling the action. The algorithm used for the update is the ``heat
bath'' algorithm, which treats the discrete set of harmonics ${\hat F}$ like a
series of coupled oscillators in thermal contact with a heat bath of fixed
temperature. The heat bath generates random fluctuations in each harmonic
which couple to all other harmonics. The fluctuations drive the system towards
a state of ``thermal''equilibrium in which the action is distributed properly.
The algorithm requires ${\cal O}[(M^2+1)N]$ operations to generate one
independent realisation, where $N$ is the number of degrees of freedom (roughly
the number of grid points for the density) and $M$
is the number of constraints. A disadvantage of this iterative approach is that
as the grid density grows and as the number of constraints increases to more
than a few, the system ``anneals'' so slowly that the algorithm becomes
prohibitively expensive and impractical. Additionally, there is no unique
way of deciding at which stage the system has annealed to the desired
equilibrium.

\medskip
\@{\bf{3.1 Hoffman-Ribak Algorithm}}
\smallskip
\@The crucial observation by Hoffman \& Ribak (1991) is that the residual field
$F({\bf x})$ has some unique properties which simplify the construction
of a realisation of a constrained field substantially. While it was already
known that the mean value of $F({\bf x})$ is independent of
the numerical values $c_i$ of the constraints $\Gamma$,

$$\langle F({\bf x})|\Gamma \rangle = \langle f({\bf x})-{\bar f}({\bf x}) |
\Gamma \rangle = \langle f({\bf x})|\Gamma \rangle - {\bar f}({\bf x}) = 0,
\eqnew$$

\@it had not been realized earlier that this is true for the complete
probability distribution ${\cal P}\left[F|\Gamma\right]$ of the residual
field $F({\bf x})$ itself (see appendix~C), i.e.

$${\cal P}\left[F|\Gamma_1\right]={\cal P}\left[F|\Gamma_2
\right]\quad\hbox{for all}\quad\Gamma_1\,,\ \Gamma_2.\eqnew$$

The observation that the statistical properties of the residual field
$F({\bf x})$ are all {\it independent} of the numerical values $c_i$ is the key
element of the Hoffman-Ribak method, rendering unnecessary a direct sampling
from the complicated action $S[F]$. A particular residual field $F({\bf x})$
can as
well have been sampled from the set of fields subject to the constraints
$\Gamma$ as from the fields belonging to some arbitrary constraint set
$\tilde \Gamma$. The residual field ${\tilde F}({\bf x})$ that is obtained by
generating an unconstrained realisation $\tilde f({\bf x})$ of the field,
and subtracting the mean field ${\bar {\tilde f}}$ of
the constraint set $\tilde \Gamma$ to which it belongs, is therefore a
correctly sampled residual field for the constraint set $\Gamma$.

These considerations lead to the following strategy for constructing a
constrained realisation of the field $f({\bf x})$, consisting of five
stages:
\item{(1)} Create a random, unconstrained, realisation $\tilde f({\bf x})$,
a homogeneous and isotropic Gaussian random field whose statistics are
determined by the power spectrum alone.
\item{(2)} Calculate for this particular realisation $\tilde f({\bf x})$ the
values
${\tilde c}_i$ of the constraints $\{C_i({\bf x})|_{{\bf x}_i}, i=1,
\dots,M\}$. These variables define a set of constraints,
$\tilde \Gamma = \{{\tilde c}_i\}$.
\item{(3)} Calculate for this ``random'' constraint set $\tilde \Gamma$ the
corresponding mean field, using

$${\bar {\tilde f}}({\bf x})\quad = \quad \langle {\tilde f}({\bf x})|
\tilde \Gamma \rangle\quad=\quad\xi_i({\bf x}) \xi_{ij}^{-1} {\tilde c}_j.
\eqnew$$

\item{(4)} Evaluate the residual field $\tilde F$ of the random
realisation:

$${\tilde F}({\bf x})={\tilde f}({\bf x})-{\bar {\tilde f}}({\bf x}).\eqnew$$

\item{} This residual field $\tilde F$ can also be
considered the residual field of a particular
realisation subject to the desired constraints, $\Gamma$.
\item{(5)} Evaluate the desired mean field ${\bar f}({\bf x})$, using
equation~(22), and add it to the residual field ${\tilde F}({\bf x})$
(eq.~34) to obtain a particular realisation of the desired constrained
Gaussian random field $f({\bf x})$:

$$f({\bf x})={\tilde f}({\bf x})+\xi_i({\bf x}) \xi_{ij}^{-1}
(c_j - {\tilde c}_j) \eqnew$$

\@The field $f({\bf x})$ constructed in this way obeys the constraints and
replaces the unconstrained field ${\tilde f}({\bf x})$. Note that there is
a one-to-one correspondence between the trial field ${\tilde f}({\bf x})$
and $f({\bf x})$. Furthermore, the ensemble of
realisations produced by the algorithm presented here properly samples the
subensemble of all realisations constrained by $\Gamma$. The algorithm is
optimal because it is exact and involves only one realisation of an
unconstrained random field and the calculation of the mean field under the
given constraints.

\medskip
\@{\bf{3.2 The practical implementation}}
\smallskip
\@Our implementation of the Hoffman-Ribak algorithm
has two important elements. Firstly, for reasons of convenience, all
necessary calculations are carried out in Fourier space. Secondly, the
constrained field $f({\bf x})$ is generated on a periodic three-dimensional
lattice of side $L$, so that $f({\bf x})$ is evaluated on $N (\propto L^3)$
gridpoints. The result can be considered to be an $N (\propto L^3)$ vector
$f=[f({\bf x}_1),\ldots,f({\bf x}_N)]$.

The central equation of the Hoffman-Ribak algorithm for generating a
constrained field realization $f({\bf x})$
is equation~(35).  We assume that, as in the case of the 18 peak
constraints (section~4), the $M$
constraints $C_i[f;{\bf x}_i]=c_i$ on the field $f({\bf x})$ are
convolutions of the field $f({\bf x})$ with some kernel $H_i({\bf x};{\bf
x}_i)$,

$$C_i[f;{\bf x}_i]=\int\,d{\bf x}\,H_i({\bf x};{\bf x}_i)\,f({\bf x})=c_i.
\eqnew$$

\@In the case of the peak constraints on the local density field (section~4.2)
the convolution kernel is a Gaussian filter function or one of
its first or second derivatives.

The Fourier transforms of the field $f({\bf x})$ and the kernel
$H_i({\bf x};{\bf x}_i)$ are defined by

$$\eqalign{f({\bf x})&=\int \d3k
\,{\hat f}({\bf k})\,e^{-i{\bf k}\cdot{\bf x}},\cr
H_i({\bf x};{\bf x}_i)&=\int \d3k
\,{\hat H}_i({\bf k})\,e^{-i{\bf k}\cdot{\bf x}}.\cr}\eqnew$$

\@Consequently, Parseval's theorem yields the following Fourier expression for
the constraint \break $C_i[f;{\bf x}_i]=c_i$,

$$C_i[f;{\bf x}_i]=\int \d3k
\,{\hat H}_i^\ast({\bf k})\,{\hat f}({\bf k})=c_i,\eqnew$$

\@The constraint's correlation function $\xi_{ij}$ can be evaluated by
using equation~(38),

$$\eqalign{\xi_{ij} \equiv \big\langle \,C_i[f;{\bf x}_i]\
C_j[f;{\bf x}_j]\,\big\rangle
&=\biggl\langle\,\int {\displaystyle d{\bf k}_1 \over
\displaystyle (2\pi)^3}\,{\hat H}_i^\ast({\bf k}_1)\,{\hat f}({\bf k}_1)\,
\int {\displaystyle d{\bf k}_2 \over
\displaystyle (2\pi)^3}\,{\hat H}_j({\bf k}_2)\,
{\hat f}^\ast({\bf k}_2)\biggr\rangle\cr
&=\int {\displaystyle d{\bf k}_1 \over \displaystyle (2\pi)^3}
{\displaystyle d{\bf k}_2 \over \displaystyle (2\pi)^3}
{\hat H}_i^\ast({\bf k}_1) {\hat H}_j({\bf k}_1)\,\,
\langle {\hat f}({\bf k}_1) {\hat f}^\ast({\bf k_2}) \rangle\,.}\eqnew$$

\@This immediately leads to the Fourier integral expression

$$\xi_{ij}=\int\,\d3k\,
{\hat H}_i^\ast({\bf k})\,{\hat H}_j({\bf k})\,P(k)\,,\eqnew$$

\@where we have used Bertschinger's definition (1992) for the spectral
density $P(k)$, modified by a factor $(2\pi)^3$ owing to our different
Fourier transform convention,

$$(2\pi)^3 P(k_1) \,\delta_D({\bf k}_1-{\bf k}_2) = \langle {\hat f}({\bf k}_1)
{\hat f}^\ast({\bf k}_2)\rangle \,,\eqnew$$

\@with $\delta_D({\bf k}_1-{\bf k}_2)$ the Dirac delta function. Once the
expression for $P(k)$ and the Fourier transform ${\hat H}_i({\bf k})$ of the
constraint kernel are known, $\xi_{ij}$ can be easily calculated from
equation~(40).

In a similar way we obtain an expression for the cross-correlation
between the field and the $i^{th}$ constraint,
$\xi_i({\bf x})$,

$$\eqalign{\xi_i({\bf x})\equiv \big\langle f({\bf x})\ C_i[f;{\bf x}_i]
\,\big\rangle
&=\biggl\langle\,
\int {\displaystyle d{\bf k}_1 \over \displaystyle (2\pi)^3}\,{\hat f}
({\bf k}_1)\,e^{-i{\bf k}_1\cdot{\bf x}} \int {\displaystyle
d{\bf k}_2 \over \displaystyle (2\pi)^3}\,{\hat H}_i({\bf k}_2)\,
{\hat f}^\ast({\bf k}_2)\,\biggr\rangle\cr
&=\int {\displaystyle d{\bf k}_1 \over \displaystyle (2\pi)^3}
{\displaystyle d{\bf k}_2 \over \displaystyle (2\pi)^3}\,
\langle {\hat f}({\bf k}_1) {\hat f}^\ast({\bf k}_2)\rangle\,
{\hat H}_i({\bf k}_2)\,e^{-i{\bf k}_1 \cdot {\bf x}}\,,}\eqnew$$

\@which in combination with the definition of the spectral density
(eq.~41) yields the expression

$$\xi_i({\bf x})=\int \d3k\,
{\hat H}_i({\bf k})\,P(k)\,e^{-i{\bf k}\cdot{\bf x}}.\eqnew$$

\@Inserting this expression into equation~(35) leads to the following
Fourier integral expression for the constrained field,

$$\eqalign{f({\bf x})&={\tilde f}({\bf x})+\xi_i({\bf x})\,\xi_{ij}^{-1}\,
(c_j-{\tilde c}_j)\cr
\cr
&=\int\,\d3k\,\biggl[{\hat{\tilde F}}({\bf k}) +
P(k)\,{\hat H}_i({\bf k})\,\xi_{ij}^{-1}\,(c_j-{\tilde c}_j)\biggr]
\,e^{-i{\bf k}\cdot {\bf x}}.\cr}\eqnew$$

\@The only element left in the calculation of the constrained
realization $f({\bf x})$ is the unconstrained field
${\tilde f}({\bf x})$. As was noted above, ${\tilde f}({\bf x})$ is most
conveniently generated in Fourier space, where its Fourier components
${\hat{\tilde F}}({\bf k})$ are mutually independent and Gaussian
distributed.

In practice the above expressions are evaluated on a three dimension grid of
$N (\propto L^3)$ gridpoints, and the corresponding Fourier integrals are
replaced
by discrete Fourier sums. Summarizing, the process of setting up a constrained
field
for a given spectrum $P(k)$ consists of four steps. Firstly, the value of the
constraint kernel is evaluated on $N$ Fourier gridpoints ${\bf k}_i$, an
${\cal O}(N)$ operation. Secondly, the matrix $\xi_{ij}$ is calculated by means
of
equation~(40), with a total computational cost proportional to ${\cal O}(M^2
N)$, after
which its inverse is determined, a ${\cal O}(M^3)$ procedure. Subsequently, the
$N$
unconstrained field components ${\hat {\tilde F}}({\bf k})$ are generated, from
which the
value of the corresponding constraint values ${\tilde c}_i$ are evaluated using
equation~(38). The computational cost of the latter is ${\cal O}(MN)$.
Finally,
the constrained field $f$ is determined from equation~(44), consisting of the
${\cal O}(M^2 N)$
evaluation of the products ${\hat H}_i({\bf k}) \xi^{-1}_{ij} (c_j-{\tilde
c}_j)$ for all
wavenumbers ${\bf k}$, followed by a Fourier transform of cost ${\cal O}(N \log
N)$.
Thus, the total cost is ${\cal O}[(M^2+\log N)N]$ (the cost of inverting the
constraints
is negligible because $N\gg M$).  Although this scaling is no better than the
${\cal O}
[(M^2+1)N]$ scaling of the iterative heat bath method (Bertschinger 1987), the
coefficient of proportionality is much smaller because no iteration is
required.

\medskip
\@{\bf {\bigone 4. The peak constraints}}
\smallskip
\@An important cosmological application for a constrained random field
algorithm is the generation of an initial density field containing one or
more peaks (or, equivalently, dips). A peak is identified as a local maximum in
the
density field that has been smoothed by some filter function or, more
generally, as the
immediate surroundings of this maximum. The choice of the
filter will depend on the specific application. The scale of the peak is
defined
to be the characteristic scale of that filter function. Depending on their
scale,
these density peaks may be the progenitors of galaxies, clusters or
superclusters. The constrained random field algorithm makes it possible
to specify the height, compactness, shape and orientation of the density
field in the immediate vicinity of the peak, while the total matter
distribution can be sculpted such that the peak is subjected to a
desired amount of net gravitational and tidal forces. In the linear
clustering regime these forces are directly proportional to the
peculiar velocity of the peak and the components of the shear at its
location.

Unlike the other constraints, the four quantities to describe
the position and scale of the peak are not imposed via the algorithm
described in the previous section. Rather, they are parameters that enter
via the kernels $H_i({\bf x};{\bf x}_i)$ (eq.~36) of each of the
constraints. In addition to its scale and location, a peak in the
smooth density field is specified by 18 constraints. The height of the
peak needs to be specified while 3 constraints are needed to ensure that
the 3 first derivatives of the smooth density field vanish at its
summit. The 6 second-order derivatives of the density field are set
by specifying the compactness, the axis ratios and the orientation of the
peak. These 10 constraints together determine the density distribution in
the immediate vicinity of the peak. The specification of the gravitational
field around the peak introduces 8 additional constraints: The 3
components of the smoothed peculiar acceleration at the location of
the peak and the 5 independent components of the traceless tidal field
tensor.

The constraints $C_i$ that we use in our peak algorithm are a combination
of one or more of the above quantities. Once a constraint has been specified
an expression for the corresponding kernel $H_i$ is derived (see eq.~36).
In the practical implementation we derive the expression for the
Fourier transform of $H_i$, ${\hat H}_i({\bf k})$. By working directly in
Fourier space we save one FFT and at the same time guarantee a higher
accuracy of the results.

After an initial phase of linear evolution in which the
Zel'dovich (1970) approximation is used, the further non-linear evolution
of the matter distribution surrounding the peak is usually followed by
N-body simulations.
It is evident that the use of the constrained random field code makes it
possible to study the formation and evolution of these objects more
systematically than possible with the conventional methods based on
unconstrained fields. Among others, this will provide considerably more
insight into the question of which physical parameters and processes
have the largest influence on the fate of an object.

\vskip 0.5cm
In the following we will drop the explicit time dependence in our notation.
The value of each of the quantities will be the value that the quantity
has when it is linearly extrapolated towards the expansion factor a (with
$a=1$ the present epoch). The treatment in the next sections will
be in comoving coordinates and wavevectors, while all spatial derivatives are
with respect to these comoving coordinates.

\medskip
\@{\bf{4.1 Peak scale and position}}
\smallskip
\@Many cosmological studies have assumed that present-day nonlinear
object like galaxies or clusters are the result of
the collapse of peaks in the primordial density fields whose height exceeds
some threshold, after having smoothed the field with a filter of a certain
shape and scale. Because many cosmological scenarios do not posses a natural
filtering scale, often an ad hoc filter has to be invoked to define the
objects. In this paper we use a Gaussian filter because of its simplicity
and smoothing properties. However, the formalism is equally
valid for any other filter, and it is trivial to modify the equations
(or our computer program) correspondingly.

Although the precise relation between the Gaussian filtering scale $R_G$ and
the characteristic mass $M_{pk}$ of a particular object in the present
universe is unclear --- indeed, the one-to-one association between
objects and density peaks is questioned by recent works (Katz, Quinn \&
Gelb 1993, Bertschinger \& Jain 1994, Van de Weygaert \& Babul 1994) ---
we can estimate a reasonable choice using a simple argument. The total
mass enclosed by a Gaussian smoothing function with filtering scale $R_G$
in a homogeneous Einstein-de Sitter universe of density ${\bar \rho}$ is

$$M_{pk}(R_G)=(2\pi)^{3/2}\,{\bar \rho} R_G^3 = 4.3718 \times 10^{12}\,
R_G^3\,h^{-1} M_{\odot},\eqnew$$

\@where $R_G$ is in units of $h^{-1}\hbox{Mpc}$. For example, if we take
for $M_{pk}$ the typical mass of the core of a cluster, $M_c=6 \times
10^{14}\,M_{\odot}$, this yields a Gaussian filter scale of $R_G \approx
4h^{-1}\,\hbox{Mpc}$. Similarly, a radius of $R_G \approx 0.6h^{-1}\,\hbox
{Mpc}$ corresponds to a mass of $\approx 10^{12}\,M_{\odot}$, comparable
to the mass of a galaxy with a luminosity equal to $L_{\ast}$
if $\Omega=1$.

\vskip 0.5truecm
The use of the filter function $W_G$ serves a twofold purpose in
our peak constraint algorithm. In addition to defining the scale of the
peaks in the density field $\rho({\bf x})$ it is also of vital
importance in the derivation of the kernels $H_i$ of each of the
constraints. The expressions for these kernels are found by using the
fact that the peaks are maxima in the filtered density field
$f_G({\bf x})$,

$$f_G({\bf x})=\int\,d{\bf y}\,f({\bf y})\,W_G({\bf y},{\bf x})\,,
\eqnew$$

\@where $f({\bf x})$ is the density contrast field,

$$f({\bf x})={\rho({\bf x})-{\bar \rho} \over {\bar \rho}}\,,\eqnew$$

\@(in this equation ${\bar \rho}$ is the average density of the Universe).
This convolution integral is equivalent to the Fourier integral

$$\eqalign{f_G({\bf x})&=\int\,\d3k\,
{\hat f}({\bf k})\,{\hat W}^*({\bf k};{\bf x})\cr
&=\int\,\d3k\,{\hat f}
({\bf k})\,{\hat W}^*({\bf k})\,e^{-i{\bf k}\cdot{\bf x}}\cr}\eqnew$$

\@where ${\hat W}({\bf k};{\bf x})$ and ${\hat W}({\bf k})$ are the
Fourier transforms of $W_G({\bf y},{\bf x})$ and $W_G({\bf x},{\bf 0})$.
In the case of a Gaussian filter,

$$W_G({\bf y},{\bf x})={1 \over ({2 \pi} R_G^2)^{3/2}}\,
\exp \left(- {|{\bf y}-{\bf x}|^2 \over 2 R_G^2}\right)\,,\eqnew$$

\@${\hat W}({\bf k};{\bf x})$ and ${\hat W}({\bf k})$ are

$${\hat W}({\bf k})=e^{-k^2 R_G^2 /2}\,\qquad\quad \hbox{and}\qquad\quad
{\hat W}({\bf k},{\bf x})={\hat W}({\bf k})\,e^{i{\bf k}\cdot{\bf x}}=
e^{-k^2 R_G^2 / 2}\,e^{i {\bf k}\cdot {\bf x}}\,.\eqnew$$

\@Note that the position ${\bf x}$ of an object causes a phase shift
${\bf k}\cdot{\bf x}$ with respect to an object that is situated at the
origin, ${\bf 0}$.

\medskip
\@{\bf{4.2 The local density field}}
\smallskip
\@Locally, the density field around a peak at position ${\bf x}_d$ can be
described by the second order Taylor expansion of the density profile
$f_G({\bf x})$ around the peak,

$$f_G({\bf x})=f_G({\bf x}_d)+{1 \over 2}\,\sum_{i,j=1}^3\,
{\partial^2 f_G \over \partial x_i \partial x_j}({\bf x}_d)\,
(x_i-x_{d,i})(x_j-x_{d,j})\,.\eqnew$$

\@In this expansion we have used the fact that the three first derivatives
of the field $f_G({\bf x})$ at the location of the local maximum,
${\bf x}_d$, are
equal to zero. The equation shows that the requirement that the smoothed
density field $f_G({\bf x})$ has a maximum of a certain height, shape,
orientation at location ${\bf x}_d$ translates into constraints on the
value of the smooth density field $f_G$ at ${\bf x}_d$, on its gradient
$\nabla f_G$ and on the second derivative tensor of the field,
$\nabla_i \nabla_j f_G$. This implies that 10 constraints are required to
fully specify the local density field around a peak.
Also note that the quadratic part of equation~(51) should be negative
definite if $f_G({\bf x}_d)$ is a maximum. Consequently,
the isodensity surfaces $f_G=F$ around the peak are triaxial ellipsoids,
whose orientation and size depends on the value of the second derivatives
of $f_G$.

\vskip 0.5truecm
The first constraint is the height of the peak, $f_G({\bf x}_d)$.
Usually it is expressed in units of the variance $\sigma_0(R_G)=
\langle f_G f_G \rangle^{1/2}$ of the smoothed density field,

$$f_G({\bf x}_d)=\nu_c \sigma_0(R_G)\,,\eqnew$$

\@which in combination with the convolution expression for $f_G$ in
equation~(46) yields the following expression,

$$\int\,\d3k\,{\hat f}
({\bf k}) \,{\hat W}^*({\bf k})\,e^{-i{\bf k}\cdot{\bf x}_d}\,=\,\nu_c
\sigma_0(R_G)\,.\eqnew$$

\@Consequently, the corresponding constraint kernel ${\hat H}_1({\bf k})$
(see eq.~38) and constraint value $c_1$ are given by

$${\hat H}_1({\bf k})={\hat W}({\bf k})\,e^{i {\bf k}\cdot
{\bf x}_d}\,,\qquad\qquad c_1 = \nu_c \sigma_0(R_G)\,.\eqnew$$

\@For reasons of clarity and convenience a compilation of the
kernels of all peak constraints is given in appendix~F.

\vskip 0.5truecm
Three additional constraints are obtained from the extremum demand that the
first order derivatives of $f_G$ should be $0$ at the peak
position ${\bf x}_d$,

$${\partial f_G \over \partial x_i} ({\bf x}_d) = 0\,,\qquad\qquad
i=1,2,3.\eqnew$$

\@The Fourier expression for the gradient $\nabla f_G ({\bf x}_d)$ is
obtained by partial differentiation of the integrand in the convolution
equation~(48),

$${\partial f_G \over \partial x_j}={\partial \over \partial x_j}
\,\int\,\d3k\,{\hat f}
({\bf k}) \,{\hat W}^*({\bf k})\,e^{-i{\bf k}\cdot{\bf x}_d}
=\int\,\d3k\,{\hat f}
({\bf k}) \,{\hat W}^*({\bf k})\,{\partial \over \partial x_j}\left(
e^{-i{\bf k}\cdot{\bf x}_d}\right)\,.\eqnew$$

\@This yields the following constraint expressions,

$$\int\,\d3k -i {\bf k} {\hat f}({\bf k})\,{\hat W}^*({\bf k})\,e^{-i {\bf k}
\cdot {\bf x}_d}\,=\,0\,.\eqnew$$

\@The corresponding kernels ${\hat H}_2({\bf k})$, ${\hat H}_3({\bf k})$
and ${\hat H}_4({\bf k})$, and the constraint values $c_2$, $c_3$ and $c_4$
are therefore

$${\hat H}_j({\bf k})= i k_l {\hat W}({\bf k})
\,e^{i {\bf k}\cdot{\bf x}_d}\,,\qquad\qquad c_j=0\,,\eqnew$$

\@where $j=2,\ldots,4$ and the corresponding $l=j-1$ (also see appendix~F).

\vskip 0.5truecm
Finally, there are six constraints that correspond to the shape,
compactness, and orientation of the density field around the peak. Because
the density field in its vicinity is ellipsoidal (see appendix~E), its shape
is fully characterized by the two axis ratios $a_{12}\equiv a_1/a_2$
and $a_{13}\equiv a_1/a_3$. The quantity that describes the compactness, or
steepness, of the density profile around a peak is the Laplacian
$\nabla^2 f_G({\bf x}_d)$. Usually this Laplacian is expressed in units of
$\sigma_2(R_G)=\langle \nabla^2 f_G \nabla^2 f_G \rangle^{1/2}$ (see
appendix~E),

$$\nabla^2 f_G ({\bf x}_d)= - x_d \sigma_2(R_G)\,.\eqnew$$

\@The minus sign in this definition of $x_d$ is introduced in order for
$x_d$ to be negative in the case of a dip and positive for a peak.
The orientation of the peak with respect to the coordinate axes is
most conveniently specified by the three Euler angles $\alpha$, $\beta$ and
$\psi$. The corresponding transformation matrix $A_{ij}$ is given by,

$$A=\pmatrix{\ \ \cos\alpha \cos\psi-\cos\beta \sin\alpha \sin\psi&
\ \ \sin\alpha \cos\psi+\cos\beta\cos\alpha\sin\psi&
\ \ \sin\beta \sin\psi\cr
-\cos\alpha \sin\psi-\cos\beta \sin\alpha \cos\psi&
-\sin\alpha \sin\psi+\cos\beta \cos\alpha \cos\psi&
-\sin\beta \cos\psi\cr
\sin\beta \sin\alpha&-\sin\beta \cos\alpha&\cos\beta\cr}.\eqnew$$

\@The above six quantities ($a_{12}$, $a_{13}$, $x_d$, $\alpha$, $\beta$ and
$\psi$) constrain the six second order derivatives of $f_G$
via the combination (see appendix~E for a derivation),

$${\partial^2 f_G \over \partial x_i \partial x_j} = - \sum_{k=1}^3 \,
\lambda_k A_{ki} A_{kj}\,,\qquad i,j=1,2,3,\eqnew$$

\@where $A_{ij}$ are the elements of the orientation matrix (eq.~60), and
the $\lambda_i$ are the eigenvalues of the matrix $-\nabla_i \nabla_j f_G$.
The values of $\lambda_i$ are obtained from the axis ratios $a_{12}$ and
$a_{13}$ of the isodensity ellipsoids around the peak, as well as
from the steepness of the density profile, $x_d$, via the relations

$$\lambda_1 = {x_d \sigma_2(R_G) \over \left(1+a_{12}^2+
a_{13}^2\right)},\qquad
\lambda_2=\lambda_1\,a_{12}^2,\qquad
\lambda_3=\lambda_1\,a_{13}^2\,,\eqnew$$

\@A Fourier expression for the second order derivatives of
$f_G({\bf x})$ is obtained by double partial differentiation of the
integrand of the convolution integral (eq.~48),

$$\int\,\d3k - k_i k_j
{\hat f}({\bf k})\,{\hat W}^*({\bf k})\,e^{-i {\bf k} \cdot {\bf x}_d}=
-\sum_{k=1}^3 \, \lambda_k A_{ki} A_{kj}\,,$$

\@so that we find

$${\hat H}_l({\bf k})=- k_i k_j {\hat W}({\bf k})\,e^{i {\bf k} \cdot {\bf
x}_d}
\,,\qquad\qquad c_l=-\sum_{k=1}^3 \, \lambda_k A_{ki} A_{kj}\,,\eqnew$$

\@for the kernels ${\hat H}_l({\bf k})$ and constraint values $c_l$, with
$l=5,\ldots,10$ and $i,j=1,\ldots,3$ (see appendix~F for the correct
numbering).

\medskip
\@{\bf{4.3 The local gravitational field}}
\smallskip
\@The peak constraints that were introduced and discussed in section~4.2
describe the density field in the immediate surroundings of the peak.
Of more fundamental importance to the dynamics of a region of space are
constraints on the gravitational potential perturbations. The local
potential perturbation $\phi({\bf x})$ is the weighted sum of all density
perturbations throughout the universe. Constraints on the local potential
therefore have immediate repercussions for the global matter distribution.
Since we wish to neglect the potential fluctuations on scales smaller than the
objects we are interested in, we consider the smoothed potential perturbation
field
$\phi_G$,

$$\phi_G({\bf x})=\int\,d{\bf y}\,\phi({\bf y})\,W_G({\bf y},{\bf
x})\,.\eqnew$$

\@In our implementation we use a Gaussian function for the filter
$W_G({\bf y},{\bf x})$, as in the case of the density field.

It is physically appealing to impose constraints on the potential $\phi$ via
constraints on its derivatives, in particular the gravitational acceleration
and the tidal field. The peculiar gravitational acceleration ${\bf g}$ at the
position ${\bf x}$ is

$${\bf g}({\bf x},t) = {1\over a} {\displaystyle d  (a {\bf v}) \over
\displaystyle dt} = - {\displaystyle 1 \over \displaystyle a} \nabla
\phi \,,\eqnew$$

\@where $a$ is the cosmological expansion factor and ${\bf v}({\bf x},t)$
the peculiar velocity of the patch of matter at physical position
${\bf r}(t)=a {\bf x}(t)$,

$${\bf v}={\displaystyle d {\bf r} \over \displaystyle dt} - H {\bf r}\,.
\eqnew$$

\@A first-order Taylor expansion of the gravitational field around the peak
shows
that the dynamical state of the patch of matter in its immediate neighbourhood,
on
scales larger than the filter scale $R_G$, is completely specified by the bulk
acceleration ${\bf g}_G({\bf x}_d)=-\nabla \phi_G/a$, the divergence
$\nabla \cdot {\bf g}_G$ and by the traceless (comoving) tidal tensor
$E_{G,ij}$,

$$g_{G,i}({\bf x})=g_{G,i}({\bf x}_d)+a\,\sum_{j=1}^3\,\left\{{\displaystyle
1 \over \displaystyle 3a}(\nabla \cdot {\bf g}_G)({\bf x}_d) \,\delta_{ij} -
E_{ij}\right\} (x_j - x_{d,j})\,,\eqnew$$

\@where $\delta_{ij}$ is the Kronecker delta and $E_{G,ij}$
is the trace-free part of $- \partial g_{G,i} / \partial r_j=
\partial^2 \phi_G/\partial r_i \partial r_j$ (note that here we choose to use
physical coordinates $r_i$, since we are dealing with physical quantities),

$$E_{G,ij} \equiv -{\displaystyle 1 \over \displaystyle 2a}
\left\{ {\displaystyle \partial g_{G,i} \over \displaystyle \partial x_i} +
{\displaystyle \partial g_{G,j} \over \displaystyle \partial x_j}\right\} +
{\displaystyle 1 \over \displaystyle 3a} (\nabla \cdot {\bf g}_G)
\,\delta_{ij}
\,=\,{\displaystyle 1 \over \displaystyle a^2} \left\{
{\displaystyle \partial^2 \phi_G
\over \displaystyle \partial x_i \partial x_j} - {\displaystyle 1 \over
\displaystyle 3} \nabla^2 \phi_G \,\delta_{ij}\right\}\,.\eqnew$$

\vskip 0.5truecm
\@The divergence $\nabla \cdot {\bf g}_G/a$ is the component of the
gravitational
field corresponding to pure radial infall into (or outflow from) the peak.
Through the Poisson equation this quantity is directly proportional to the
local
density perturbation $f_G({\bf x})$,

$${\displaystyle \nabla \cdot {\bf g}_G \over \displaystyle a} =
-{\displaystyle 1 \over \displaystyle a^2}
\nabla^2 \phi_G= - {\displaystyle 3 \over \displaystyle 2} \Omega H^2\,
\, f_G({\bf x})\,.\eqnew$$

\vskip 0.5truecm
\@The expression for the constraint on $\nabla \cdot {\bf g}_G/a$ is therefore
equivalent to equation~(53), except for the proportionality constant
$3\Omega H^2/2$ in both constraint kernel ${\hat H}_j$ and value $c_j$. From
the
above equation we can also easily infer the relation between the Fourier
components
${\hat \phi}_G({\bf k})$,

$$\phi_G({\bf x})=\int\,\d3k\,
{\hat \phi}_G({\bf k})\,e^{-i{\bf k}\cdot{\bf x}}\,,\eqnew$$

\@and the Fourier components ${\hat f}({\bf k})$ of the density field,

$${\hat \phi}_G({\bf k})=
-{\displaystyle 3 \over \displaystyle 2} \Omega H^2 a^2\,{\displaystyle 1 \over
\displaystyle k^2}\,{\hat f}({\bf k}) {\hat W}^{\ast}({\bf k})\,.\eqnew$$

\vskip 0.5truecm
\@The first 3 constraints on the gravitational field therefore concern the
peculiar gravitational acceleration at the position of the peak itself,
${\bf g}_G({\bf x}_d)$. It is useful to specify it in units of the dispersion
of
the gravitational acceleration of peaks,
$\sigma_{g,pk}(R_G)=\langle{\bf g}_{G,pk}\cdot {\bf g}_{G,pk}\rangle$,

$$g_{G,l}({\bf x}_d)={\tilde g}_{l}\,\sigma_{g,pk}(R_G)\,,\qquad
l=1,\ldots,3\,.\eqnew$$

\@The dispersion of the peak accelerations is less than the overall dispersion
$\sigma_{g}$ of the acceleration in the field. This lowering of the
acceleration
of peaks compared with that of field points is caused by the extra acceleration
associated with the infall of field points onto the peaks. We can infer that
(see section 4.4, eqns.~101 and 106)

$$\sigma_{g,pk}={\tilde
\sigma}_g\equiv\sigma_{g}\,\sqrt{1-\gamma_v^2}\,,\eqnew$$

\@where $\sigma_g(R_G)$ and $\gamma_v$ are given by

$$\sigma_{g}(R_G)={\displaystyle 3 \over \displaystyle 2}\Omega H^2 \,
\sigma_{-1}(R_G)\qquad \hbox{and} \quad
\gamma_v\equiv{\displaystyle \sigma_0^2 \over \displaystyle \sigma_{-1}
\sigma_1}
\,,\eqnew$$

\@with $\sigma_j(R_G)$ the spectral moments,

$$\sigma_j^2(R_G) \equiv \int\,\d3k\,
P({\bf k})\,{\hat W}({\bf k})\,k^{2j}\,.\eqnew$$

\@The Fourier expressions for the 3 components of the bulk peculiar
acceleration ${\bf g}_G({\bf x}_d)$ can be derived from equation~(65) and
(70),

$$g_{G,l}({\bf x}_d)=-{\displaystyle 1 \over \displaystyle a}
{\partial \phi_G \over \partial x_l}=
\int\,\d3k\,{\displaystyle 1 \over \displaystyle a} \, i k_l
{\hat \phi}_G({\bf k})\,e^{-i{\bf k}\cdot{\bf x}}\,.\eqnew$$

\@Inserting equation~(71) and (72) leads to the constraint equations,

$$\int\,\d3k\,{\hat f}({\bf k})
\left\{-{\displaystyle 3\over \displaystyle 2} \Omega H^2
\,{\displaystyle i k_l \over \displaystyle k^2} {\hat W}^{\ast}({\bf k})
\,e^{-i{\bf k}\cdot{\bf x}_d}\right\}
={\tilde g}_{l}\,{\displaystyle 3\over \displaystyle 2} \Omega H^2
\,\sqrt{1-\gamma_v^2}\,\sigma_{-1}(R_G)\,,\eqnew$$

\@From this we find the corresponding constraint kernels ${\hat H}_{11}({\bf
k})$,
${\hat H}_{12}({\bf k})$ and ${\hat H}_{13}({\bf k})$, and the constraint
values $c_{11}$, $c_{12}$ and $c_{13}$,

$${\hat H}_j({\bf k})={\displaystyle 3\over \displaystyle 2}\Omega H^2\,
\,{\displaystyle i k_l \over
\displaystyle k^2}\,{\hat W}({\bf k})\,e^{i {\bf k} \cdot {\bf x}_d}\,,\qquad
c_j={\tilde g}_{l}\,{\displaystyle 3\over \displaystyle 2} \Omega H^2\,
\sqrt{1-\gamma_v^2}\,\,\sigma_{-1}(R_G),\eqnew$$

\@with $j=11,\ldots,13$ and $l=1,\ldots,3$ (see appendix~F). Evidently, instead
of
specifying the constraint values $c_j$ as ${\tilde g}_{l}$ it is also possible
to
do this directly in the appropriate physical units ({\it e.g.} km/s$^2$).

\vskip 0.5truecm
Five additional constraints are needed to characterize the tidal field around
the
peak. This field is described by the traceless (comoving) tidal tensor
$E_{G,ij}$
(eq.~68). Within an arbitrary system of reference this tidal tensor is most
conveniently expressed in terms of its eigenvalues and units vectors,

$$E_{G,ij}={\displaystyle 1 \over \displaystyle a^2}\left\{
{\displaystyle \partial^2 \phi_G
\over \displaystyle \partial x_i \partial x_j} - {\displaystyle 1 \over
\displaystyle 3} \nabla^2 \phi_G \,\delta_{ij}\right\}=
\sum_{k=1}^3 \, {\cal E}_k T_{ki} T_{kj}\,,\qquad i,j=1,2,3.\eqnew$$

\@The elements of the matrix $T_{kl}$ are the components of the various
eigenvectors of the
tidal tensor, whose directions are characterized by the 3 Euler angles
$\alpha_E$, $\beta_E$, and
$\psi_E$ ($T_{kl}$ are given by equation~(60), with $\alpha_E$, $\beta_E$ and
$\psi_E$
replacing $\alpha$, $\beta$ and $\psi$). In an initial random density field
there is a strong
correlation between the tidal tensor and the mass tensor $\zeta_{ij}=\nabla_i
\nabla_i f$
(see section 4.4, eq.~106). In the case of peaks this translates into a strong
tendency of the
tidal tensor to align itself along the principal axes of the mass ellipsoid. In
the specification
of the initial tidal field it is therefore often useful, and physically
sensible, to
express its elements with respect to the reference system defined by these
axes. We denote the
corresponding transformation matrix by ${\tilde T}_{kl}$, which is defined
through equation~(60)
by the 3 corresponding Euler angles ${\tilde \alpha}_E$, ${\tilde \beta}_E$ and
${\tilde \psi}_E$. If the orientation of the peak itself with respect to an
arbitrary
reference system is specified by the transformation matrix
$A_{kl}(\alpha,\beta,\psi)$
(see eq.~60), then the tidal field's transformation matrix $T$ within the same
system is
the matrix product of ${\tilde T}$ with $A$,

$$T_{ki} = \sum_{m=1}^3\,{\tilde T}_{km} A_{mi}\,,\eqnew$$

The magnitude of the tidal field in the directions of the principal axes of the
tidal tensor
is given by the eigenvalues ${\cal E}_1$, ${\cal E}_2$ and ${\cal E}_3$.
Because $E_{G,ij}$ is traceless, {\it i.e.} $\sum {\cal E}_k = 0$, it is
sufficient to
specify two eigenvalues. To get an idea of the right order of magnitude it is
usually useful to
specify ${\cal E}_k$ in units of $\sigma_E$, the
dispersion of the off-diagonal elements of the tidal tensor $E_{G,ij}$ (see
section 4.4,
eq.~101),

$$\sigma_{E}(R_G)={\displaystyle 3 \over \displaystyle 2} \Omega
H^2\,\sigma_0(R_G)\,
\sqrt{\displaystyle 1 - \gamma^2 \over \displaystyle 15}\qquad \hbox{with}
\quad
\gamma\equiv{\displaystyle \sigma_1^2 \over \displaystyle \sigma_{0} \sigma_2}
\,,\eqnew$$

\@so that ${\cal E}_k={\tilde {\cal E}}_k\,\sigma_E(R_G)$. An elegant and
convenient
parameterization of the diagonalized $E_{G,ij}$ in terms of two quantities
$\epsilon$ and $\varpi$ was introduced by Bertschinger \& Jain (1994),

$$E_{G,ij}=\hbox{diag} \,\left[{\cal E}_1,{\cal E}_2,{\cal E}_3\right]\equiv
\Omega H^2 \,\epsilon\,
(1+f_G) \, Q_{ij}(\varpi)\,,\eqnew$$

\@with the one-parameter traceless matrix $Q_{ij}$ defined by

$$Q_{ij}(\varpi) \equiv \hbox{diag}\,\left[{\cal Q}_1,{\cal Q}_2,{\cal
Q}_3\right]
\equiv \hbox{diag}\,\left[ \cos\left({\displaystyle \varpi + 2\pi \over
\displaystyle 3}\right)\,,\,\cos\left({\displaystyle \varpi - 2\pi \over
\displaystyle 3}\right)\,,\,\cos\left({\displaystyle \varpi \over \displaystyle
3}
\right)\,\right]\,.\eqnew$$

\@This matrix representation turns out to be useful when considering the
Lagrangian equations of motion of a patch of matter (Bertschinger \& Jain
1994).
It is particularly convenient because all possible eigenvalues of $E_{G,ij}$
are
obtained by $qQ_{ij}(\alpha)$, with $q \in [0,\infty)$ determining the
magnitude
of the tidal field, and $\alpha \in [0,\pi]$ the relative strength of the tidal
field
along the three principal axes. The 5 constraints, for $E_{G,11}$, $E_{G,22}$,
$E_{G,12}$, $E_{G,13}$ and $E_{G,23}$, therefore have the form

$$E_{G,ij}={\tilde \epsilon}\,\sigma_{E}(R_G)\,\sum_{k=1}^3\,
{\cal Q}_k(\varpi) T_{ki} T_{kj}\,,\eqnew$$

\@with $(i,j)=(1,1),(2,2),(1,2),(1,3)$ and $(2,3)$. Note that here we have
expressed
$\epsilon$ in units of $\sigma_E$, i.e. ${\tilde {\cal E}}={\tilde \epsilon}\,
{\cal Q}(\varpi)$. In addition, we have assumed that the fluctuations are
linear, so that
the factor $\epsilon f_G$ can be neglected. For the generation of initial
conditions, our
primary interest, this assumption is not a serious restriction.

The Fourier components ${\hat E}_{G,ij}({\bf k})$ of the tidal tensor,

$$E_{G,ij}=\int\,\d3k\,
{\hat E}_{G,ij}({\bf k})\,e^{-i{\bf k}\cdot{\bf x}_d}\,,\eqnew$$

\@can be easily found from the definition of $E_{G,ij}$ in equation~(68) and
subsequent differentiation and insertion of equation~(71),

$${\hat E}_{G,ij}({\bf k})=
{\displaystyle 3 \over \displaystyle 2}\Omega H^2 \,\left({\displaystyle
k_i k_j \over \displaystyle k^2}-{\displaystyle 1 \over \displaystyle 3}
\delta_{ij}\right)\,{\hat W}^*({\bf k})\,{\hat f}({\bf k})\,.\eqnew$$

\@This leads to the following tidal field constraint expressions:

$$\int\,\d3k\,
\left\{{\displaystyle 3 \over \displaystyle 2}\Omega H^2 \,\left({\displaystyle
k_i k_j \over \displaystyle k^2}-{\displaystyle 1 \over \displaystyle 3}
\delta_{ij}\right)\,{\hat W}^*({\bf k})\,e^{-i{\bf k}\cdot{\bf x}_d}\right\}
{\hat f}({\bf k})\,\,={\tilde \epsilon}\,\sigma_E(R_G)\,\sum_{k=1}^3\,
{\cal Q}_k(\varpi) T_{ki} T_{kj}\,,\eqnew$$

\@which yields the corresponding 5 constraint kernels and values,

$${\hat H}_l({\bf k})=
{\displaystyle 3 \over \displaystyle 2}\Omega H^2 \,\left({\displaystyle
k_i k_j \over \displaystyle k^2}-{\displaystyle 1 \over \displaystyle 3}
\delta_{ij}\right)\,{\hat W}({\bf k})\,e^{i{\bf k}\cdot{\bf x}_d}\,,\qquad
c_l={\tilde \epsilon}\,{\displaystyle 3 \over \displaystyle 2} \Omega H^2\,
\sigma_0(R_G)\,\sqrt{\displaystyle 1 - \gamma^2 \over \displaystyle 15}
\,\sum_{k=1}^3\,{\cal Q}_k(\varpi) T_{ki} T_{kj}\,,\eqnew$$

\@with $l=13,\ldots,18$ and $(i,j)=(1,1),(2,2),(1,2),(1,3)$ and $(2,3)$.
Alternatively,
instead of expressing the tidal constraints via the two quantities $\epsilon$
and $\varpi$
(or ${\cal E}$) and the 3 Euler angles $\alpha_E$, $\beta_E$ and $\psi_E$, we
can evidently
specify the values for $E_{G,ij}$ directly, either in corresponding
physical units or in units of $\sigma_E(R_G)$,

$$E_{G,ij}={\tilde \varepsilon}_{ij}\,{\displaystyle 3 \over \displaystyle 2}
\Omega H^2\,
\sigma_0(R_G)\,\sqrt{\displaystyle 1 - \gamma^2 \over \displaystyle
15}\,.\eqnew$$

\vskip 0.5truecm
In the linear regime an analogous, and for some more familiar, way of
describing
the dynamics of a patch of matter is in terms of the peculiar velocity. This is
possible because for growing mode linear perturbations the peculiar velocity
${\bf v}$ is directly proportional to the peculiar gravitational acceleration
${\bf g}$ (Peebles 1980),

$${\bf v}({\bf x},t)={\displaystyle 2 {\cal F}(\Omega) \over \displaystyle 3 H
\Omega}\, {\bf g}({\bf x},t)\,,\eqnew$$

\@where ${\cal F}(\Omega) \approx \Omega^{0.6}$. It is convenient to write the
smoothed peculiar velocity ${\bf v}_G$ around the position ${\bf x}_d$ of the
peak
in terms of the bulk motion ${\bf v}_G({\bf x}_d)$, the divergence
$\nabla \cdot {\bf v}_G/a$, the shear $\sigma_{ij}$ and the
vorticity $\omega_{ij}$,

$$v_{G,i}({\bf x})=v_{G,i}({\bf x}_d)+a \sum_{j=1}^3 \left\{{\displaystyle 1
\over
\displaystyle 3a} (\nabla \cdot {\bf v}_G)({\bf x}_d)\,\delta_{ij} +
\sigma_{ij} ({\bf x}_d) + \omega_{ij} ({\bf x}_d)\right\}\,(x_j-x_{d,j})\,.
\eqnew$$

\@The shear is the trace-free symmetric part of ${\partial v_{G,i} / \partial
r_j}$,

$$\sigma_{ij}={\displaystyle 1 \over \displaystyle 2 a}
\left\{ {\partial v_{G,i}
\over \partial x_j}+{\partial v_{G ,j} \over \partial x_i}\right\} -
{\displaystyle 1 \over \displaystyle 3 a} (\nabla \cdot {\bf
v}_G)\,\delta_{ij}\,,\eqnew$$

\@while the vorticity $\omega_{ij}$ is the antisymmetric part,

$$\omega_{ij}={\displaystyle 1 \over \displaystyle 2 a}
\left\{{\partial v_{G,i} \over \partial x_j}-{\partial v_{G,j} \over
\partial x_i}\right\}\,.\eqnew$$

\@Because $\omega_{ij}$ does not have a gravitational origin, it is an
irrelevant
quantity as far as constraints on the density perturbation field are concerned.
Moroever, it can be shown to remain zero whenever there is no primordial
vorticity
(Peebles 1980). From this we can infer that constraints on the peak velocity
${\bf v}_G({\bf x}_d)$ are therefore equivalent to equation~(77), except that
the
factor $3\Omega H^2 a/2$ in both the constraint kernel ${\hat H}_j$ and
constraint
value $c_j$ has to be changed into $H a {\cal F}(\Omega)$. Also, the constraint
on
the divergence $\nabla \cdot {\bf v}_G/a$ is equivalent to the constraint on
$f_G({\bf x})$ (eq.~53), except for a factor $H {\cal F}(\Omega)$.
Finally, we see that the relation between the shear $\sigma_{G,ij}$ and
$E_{G,ij}$
is

$$E_{ij}=-{\displaystyle 3 \over \displaystyle 2} \Omega H^2
{\displaystyle \sigma_{ij} \over \displaystyle H {\cal F}(\Omega)}\,.\eqnew$$
\@We should, however, not fail to appreciate that in the nonlinear regime the
simple relation between ${\bf v}$ and ${\bf g}$ breaks down. In that case
the same basic physical relationships between the acceleration ${\bf g}$, the
potential $\phi$ and the density $\rho$ remain valid. As this is not true for
the velocity ${\bf v}$, it is fundamentally preferrable to impose constraints
on the
gravitational field instead of the velocity field.

\medskip
\@{\bf{4.4 Probability of peak constraints}}
\smallskip
\@In principle, in the case of a Gaussian field any set of numerical values for
the
18 peak constraints per peak is possible, regardless of how small the
probability of the
occurrence of such peaks would be. This is a consequence of the ability of the
Hoffman-Ribak constrained random field method to generate realizations for {\it
any
arbitrary} set of values for the imposed constraints. In order to prevent the
generation
of unlikely circumstances it is therefore necessary to control or have an
estimate of the
likelihood of the constraints. The corresponding probability distribution of
the constraints
is given by equation~(15). A good measure of the likelihood can be obtained by
calculating the
$\chi^2$,

$$ \chi^2 \ \ = \ \ \sum_{i,j=1}^{M}\,\,C_i\,({\bf Q}^{-1})_{ij}\,C_j \ \ = \ \
c_i\,\xi_{ij}^{-1} c_j\,.\eqnew$$

\@The probability that for this constraint set $\chi^2$ has this value or
higher can then be
directly calculated from $\Gamma_Q(M/2,\chi^2/2)$, where $\Gamma_Q$ is the
incomplete gamma
function.
As a rule of a thumb, the constraint set can be considered to represent
manifest unlikely
conditions, if the $\chi^2$ per degree of freedom, $\tilde\chi^2 \equiv
\chi^2/M$, differs
significantly from unity. Note that the computational cost of evaluating
$\tilde\chi^2$ is
negligible as the inverse of the constraint-constraint correlation matrix
$\xi_{ij}=\langle
C_i C_j\rangle$ has already been calculated as part of the construction
procedure
(eq.~40 and 44).

A full expression for $\chi^2$ or, even better, the full probability
distribution in terms of
the 18 constraint quantities can be obtained by evaluating the expression in
equation~(15),
following the treatment presented in Appendix A of BBKS. Following the
discussion in
the previous sections the density and gravity field in and around an arbitrary
point ${\bf x}$
in a Gaussian random density field $f({\bf x})$ can be characterized by 18
parameters

$$\Upsilon = \{ \nu, \eta_1, \eta_2, \eta_3, \zeta_1, \zeta_2, \zeta_3,
\zeta_4, \zeta_5,
\zeta_6, g_1, g_2, g_3, E_1, E_2, E_4, E_5, E_6\}\,,\eqnew$$

\@with $f=\nu \sigma_0$ the value of the field at ${\bf x}$, $\nabla_i f =
\eta_i$ the first
derivatives of the field, and $\zeta_A$ the six independent components of the
tensor
$\zeta_{ij}=\nabla_i \nabla_j f$ (where $A=1,2,3,4,5,6$ refer to the
$ij=11,22,33,12,13,23$
components of the tensor). In addition, $g_i=-\nabla_i \phi$ is the peculiar
gravitational
acceleration, while $E_A$ are the five independent components of the traceless
tidal
tensor $E_{ij}=\nabla_i \nabla_j \phi - {1 \over 3} \nabla^2 \phi\,\delta_{ij}$
(with
$A=1,2,3,4,5,6$ referring to the $ij=11,22,33,12,13,23$ components of
$E_{ij}$).

The probability ${\cal P}(\Upsilon)$ that at the position ${\bf x}$ the field
has the specified
values for these 18 quantities is specified by a joint Gaussian
probability distribution, for which a reasonably insightful expression can be
found by reducing the corresponding $18 \times 18$ covariance matrix ${\bf
Q}=\langle y_i
y_j\rangle$ into a block diagonal matrix of 9 $2 \times 2$ blocks. This is
achieved by
transformation of the set of variables $\{\zeta_1,\zeta_2,\zeta_3,E_1,E_2\}$
into a new set
$\{x,y,z,E_y,E_z\}$,

$$\eqalign{ x = - {\displaystyle \zeta_1 + \zeta_2 +\zeta_3 \over \displaystyle
\sigma_2}\,,
\qquad y &=-{\displaystyle \zeta_1 - \zeta_3 \over \displaystyle 2 \sigma_2}\,,
\qquad z = -{\displaystyle \zeta_1 - 2 \zeta_2 + \zeta_3 \over \displaystyle 2
\sigma_2}\,,\cr
\ \ \ \cr
E_y&={\displaystyle E_1 - E_3 \over \displaystyle 2}\,,\qquad
E_z={\displaystyle
E_1 - 2 E_2 + E_3 \over \displaystyle 2}\,.\cr}\eqnew$$

\@${\cal P}(\Upsilon)$ is then given by

$${\cal P}(\Upsilon)\quad = \quad A\,e^{-Q/2}\
d\nu\,d^3 \eta\,dx dy dz\,d\zeta_4 d\zeta_5 d\zeta_6\,d^3 g\,dE_y dE_z\,dE_4
dE_5 dE_6\,,\eqnew$$

\@with

$$A\ \ = \ \ {\displaystyle 3^6\,5^5 \over \displaystyle
1024\,\pi^9\,(1-\gamma^2)^3\,
(1-\gamma_v^2)^{3/2}\,({3 \over 2} \Omega
H^2)^4\,\sigma_{-1}^3\,\sigma_0\,\sigma_1^3\,
\sigma_2^3}\ ,\eqnew$$

\@and

$$\eqalign{Q=\quad &\sum_{i,j=1}^{18} y_i\,({\bf Q}^{-1})_{ij} y_j \quad = \cr
=\quad &\nu^2 \,+ \,{\displaystyle (x-x_\ast)^2 \over \displaystyle 1-\gamma^2}
\,+\,
15y^2 \,+\,5z^2 \,+\, {\displaystyle 3 {\vec \eta} \cdot {\vec \eta} \over
\displaystyle
\sigma_1^2} \,+\, \sum_{A=4}^6 {\displaystyle 15 \zeta_A^2 \over \sigma_2^2}
\,+ \cr
&\,{\displaystyle 3 ({\vec g}-{\vec g}_\ast)^2 \over \displaystyle
\tilde\sigma_g^2} \,+\,
{\displaystyle (E_y -E_y^\ast)^2 \over \displaystyle \sigma_E^2} \,+\,
{\displaystyle (E-E_z^\ast)^2 \over \displaystyle 3 \sigma_E^2} \,+\,
\sum_{A=4}^6
{\displaystyle (E_A-E_A^\ast)^2 \over \displaystyle \sigma_E^2}\,,\cr}\eqnew$$

\@where $\tilde\sigma_g$ and $\sigma_E$ are defined by

$$\tilde\sigma_g \equiv ({3 \over 2} \Omega
H^2)\,\sigma_{-1}\,\sqrt{1-\gamma_v^2}\,,\qquad\qquad
\qquad\sigma_E \equiv ({3 \over 2} \Omega H^2)\,\sigma_0 \sqrt{\displaystyle 1
- \gamma^2 \over
\displaystyle 15}\,,\eqnew$$

\@while the various coupling quantities $x_\ast$, ${\vec g}_\ast$, $E_y^\ast$,
$E_z^\ast$
and $E_A^\ast$ are defined by

$$\eqalign{x_\ast &= \gamma\,\nu\,,\cr
{\vec g}_\ast &= \gamma_v\,({3 \over 2} \Omega H^2)\, {\displaystyle
\sigma_{-1} \over
\displaystyle \sigma_1}\,{\vec \eta}\cr
E_y^\ast&=\gamma\,y\,({3 \over 2} \Omega H^2)\,\sigma_0\,,\qquad
E_z^\ast=\gamma\,z\,({3 \over 2} \Omega H^2)\,\sigma_0\,,\qquad\cr
E_A^\ast&=\gamma\,({3 \over 2} \Omega H^2)\,{\displaystyle \sigma_0 \over
\displaystyle
\sigma_2}\,\zeta_A\,,\qquad A=4,5,6\,.\cr}\eqnew$$

\@(for the definitions of $\gamma$, $\gamma_v$ and the various $\sigma_j$ see
eqns.~81,
74 and 75). In the case of a peak we can further reduce the expression for $Q$.
Evidently,
$\vec \eta = 0$. In addition, we can use the
fact that $Q$ should be independent of the orientation of the mass ellipsoid
around the
peak, expressed by its Euler angles $\alpha$, $\beta$ and $\psi$. We can
therefore
discard the orientation term $\sum 15 \zeta_A^2 /\sigma_2^2$ and redefine $x$,
$y$ and
$z$ in terms of the eigenvalues $\lambda_i$ of $(-\zeta_{ij})$, whose relation
to the axis
ratios of the ellipsoid are given in equation~(62),

$$x = {\displaystyle \lambda_1 + \lambda_2 +\lambda_3 \over \displaystyle
\sigma_2}\,,
\qquad y ={\displaystyle \lambda_1 - \lambda_3 \over \displaystyle 2
\sigma_2}\,,
\qquad z ={\displaystyle \lambda_1 - 2 \lambda_2 + \lambda_3 \over
\displaystyle 2 \sigma_2}
\,,\eqnew$$

\@Furthermore, we restrict the ordering of the eigenvalues to $\lambda_1 \geq
\lambda_2
\geq  \lambda_3 > 0$. The condition that $\lambda_3 > 0$ is equivalent to the
demand that
$\zeta_{ij}$ has to be negative definite, which, together with the constraints
$\nabla_i f$,
is necessary and sufficient to have a local density maximum, a peak. We then
find for the
complete probability ${\cal P}(\Upsilon)$ that at an arbitrary position ${\bf
x}$ there is
a peak with a height
$f=\nu \sigma_0$, a shape characterized by the parameters $x$, $y$ and $z$, an
orientation
specified by the Euler angles $\alpha$, $\beta$ and $\psi$, an acceleration
${\vec g}$,
and a tidal field described by the parameters $E_y$, $E_z$, $E_4$, $E_5$ and
$E_6$, or,
rather, that there is a peak with these parameters in the specific
infinitesimal ranges
around these values,

$$\eqalign{&{\cal P}(\nu,x,y,z,\alpha,\beta,\psi,{\vec
g},E_y,E_z,E_4,E_5,E_6)\cr
&\qquad=\quad \tilde A\,|y(y^2-z^2)|\,\sin \beta\,e^{-\tilde
Q/2}\,d\nu\,d^3\eta\,dx dy dz\,
d\alpha d\beta d\psi\,d{\vec g}\,dE_y dE_z\,dE_4 dE_5 dE_6\,.\cr}
\eqnew$$

\@In this equation the constant $\tilde A$ is

$$\tilde A \ \ =\ \ {\displaystyle 3^7\,5^5 \over \displaystyle
256\,\pi^9\,(1-\gamma^2)^3\,
(1-\gamma_v^2)^{3/2}\,({3 \over 2} \Omega
H^2)^4\,\sigma_{-1}^3\,\sigma_0\,\sigma_1^3}
\ ,\eqnew$$

\@and

$${\tilde Q}\quad=\quad \nu^2 \,+ \,{\displaystyle (x-x_\ast)^2 \over
\displaystyle 1-\gamma^2}
\,+\, 5(3y^2+z^2) \,+\,{\displaystyle 3 {\vec g}^2 \over \displaystyle
\tilde\sigma_g^2} \,+\, {\displaystyle (E_y -E_y^\ast)^2 \over \displaystyle
\sigma_E^2} \,+ \,
{\displaystyle (E_z-E_z^\ast)^2 \over \displaystyle 3 \sigma_E^2} \,+\,
\sum_{A=4}^6
{\displaystyle E_A^2 \over \displaystyle \sigma_E^2}\,,\eqnew$$

\@where $E_y$, $E_z$, $E_4$, $E_5$ and $E_6$ are specified with respect to the
principal axis of the mass ellipsoid (see section 4.3, eq.~80 for the
appropriate
transformations). In particular, this implies that $E_A^\ast=0$ for
$A=4,5,6$ (eq.~102). Also note that equation~(100) is therefore an expression
of the fact that
in an initial random density field the tidal field has a strong preference to
align itself along
the principal axes of the the mass tensor $\zeta_{ij}$.  In particular, for a
peak this
implies that the strongest tidal force tends to be directed along its smallest
axis
(the one with the highest eigenvalue $\lambda_i$). As it explicitly takes into
account this
strong correlation between the initial mass quadrupole and the tidal field at
the position
of the peak, the reference system defined by the mass ellipsoid is therefore
the most
natural one to specify the initial tidal forces. The resulting expression for
${\tilde Q}$
in the above equation is essentially the one for the $\chi^2$ of the imposed
constraints,
once it is scaled to the appropriate filter and filter scale $R_G$ by means of
$\gamma(R_G)$, $\gamma_v(R_G)$ and the various spectral moments
$\sigma_j(R_G)$.

The probability distribution in equation~(104) is the one for having, at some
arbitrary field position, a peak with the required physical parameters.  Often,
however, we are
more interested in the more specific question what the probability ${\cal
P}_{pk}$ is that a
peak at an arbitrary position has these imposed constrained properties. To
evaluate this we need
to determine the (comoving) number density of peaks with the constrained
parameters,
which can be done following the prescription in BBKS. To obtain ${\cal P}_{pk}$
this specific
number density has to be divided by the total comoving number density of peaks,
$n_{pk}$, whose
value is (see BBKS),

$$n_{pk}\ \ =\ \ {\displaystyle 29 - 6\sqrt{6} \over \displaystyle 5^{3/2} 2
(2\pi)^2\, R_\ast^3}
\ \ =\ \ 0.016\,R_\ast^{-3}\,,\qquad\qquad\hbox{with}\qquad R_\ast \equiv
\sqrt{3}\,
{\displaystyle \sigma_1 \over \displaystyle \sigma_2}\,.\eqnew$$

\@From this we can  derive the probability that a peak has a height $\nu
\sigma_0$, shape
parameters $x$, $y$ and $z$, an acceleration ${\vec g}$ and tidal tensor
components
$E_y$, $E_y$, $E_4$, $E_5$ and $E_6$. Since the orientation of the peak is here
a less relevant
quantity we integrate over the Euler angles $\alpha$, $\beta$ and $\psi$. This
can be done
without further complications since ${\tilde Q}$ is independent of these Euler
angles. Note that this automatically implies that the tidal tensor components
are specified with
respect to the principal axes of the mass ellipsoid. We then obtain the
following
expression for ${\cal P}_{pk}$,

$$\eqalign{&P_{pk}(\nu,x,y,z,{\vec g},E_y,E_z,E_4,E_5,E_6)\,d\nu\,dx dy
dz\,d{\vec g}\,
dE_y dE_z\,dE_4 dE_5 dE_6\cr
&\qquad=\quad {\tilde B}\,\Theta(x,y,z)\,F(x,y,z)\,e^{-{\tilde Q}/2}\,d\nu\,dx
dy dz\,
d{\vec g}\,dE_y dE_z\,dE_4 dE_5 dE_6\,.\cr}\eqnew$$

\@In this expression ${\tilde Q}$ is given by equation~(106), while

$$F(x,y,z)\ =\ \,y\,(y^2-z^2)\,(x-2z)\,[(x+z)^2-(3y)^2]\,.\eqnew$$

\@In addition, the function $\Theta(x,y,z)$ is defined such that its value is 1
when the peak
constraints in the $(x,y,z)$ domain are satisfied, and 0 otherwise. These
constraints are
that $y \geq z \geq -y, \ y \geq 0$, to obtain the correct ordering of the
eigenvalues
$\lambda_i$ of $\zeta_{ij}$, and $(x+z-3y) > 0$ so that the smallest eigenvalue
$\lambda_3$ is
positive and we indeed have a peak. The constant ${\tilde B}$ is given by

$${\tilde B} \ \ = \ \ {\displaystyle 5^{13/2}\,3^{17/2} \over \displaystyle
16\,\pi^5\,(29-6\sqrt{6})\,(1-\gamma^2)^3\,(1-\gamma_v^2)^{3/2}\,({3 \over 2}
\Omega H^2)^4\,\sigma_{-1}^3\,\sigma_0}\,.\eqnew$$

\@This can be easily extended to other conditional peak probabilities, e.g. the
chance that
a peak of height $\nu \sigma_0$ has the required parameters. However,
calculating the
involved expressions quickly becomes a very elaborate procedure.

In the above we have mainly concentrated on the probability of constraints
imposed on one
particular peak, with the intention of providing insight into how the different
constraints
interrelate and to get an idea of the expected order of magnitude of each of
the constraints.
However, as we have seen earlier, our code allows to provide constraints on
many
different peaks, at different positions and scales. Giving analytical
expressions for such
constraints would quickly become a cumbersome and elaborate affair, due to the
introduction of spatial correlations in the random field. However, via
equation~(95) the
numerical value of $\chi^2$ for these constraints can be easily computed,
providing a good
idea of their likelihood.

\medskip
\@{\bf {\bigone 5. Realizations and Applications}}
\smallskip
\@The formalism developed in the previous sections allows the generation of a
large
variety of initial conditions. In this section we will visualize the procedure
by
providing some practical examples. The versatile and non-local nature of the
formalism has already
been emphasized in figure~1 (section~2), illustrating the construction of a
density field
that is constrained to have two peaks of a different scale, a different shape,
and at
different positions. Although for the construction of the density field around
one central peak a
few other equally or even more efficient methods have been developed (Binney \&
Quinn 1991,
Bond \& Meyers 1993), mainly based on a multipole expansion of the field, their
efficiency
breaks down if the constraints are imposed at more than one position.

\medskip
\@{\bf{5.1 Field-constraint correlations}}
\smallskip
\@At the core of our construction procedure is the superposition of a mean
field ${\bar f}$
and a properly sampled residual field $F$ (see eq.~26). The mean field ${\bar
f}$
is effectively the superposition of the field-constraint correlation fields
$\xi_i({\bf x})$
(eq.~22), each weighted by a factor $\sum\nolimits_j \xi_{ij}^{-1} c_j$.

Figure~3 shows the mean field and five of the composite field-constraint
correlation fields
for the set of peak constraints described below. These realizations are
generated in a periodic
$100h^{-1}\,\hbox{Mpc}$ box. As for figure~1, the power spectrum of the random
field
fluctuations is the standard cold dark matter spectrum of Davis et al.  (1985)
with
$\Omega_{\hbox{CDM}}=1.0$ and $h=0.5$ --- normalized such that
$\sigma(8h^{-1}\,\hbox{Mpc})
=1.0$ at $a=1$, the present epoch (Davis \& Peebles 1983). The panels in
figure~3 contain
contourmaps of the density and correlation values in a $5h^{-1}\,\hbox{Mpc}$
slice centered halfway in the simulation box, each of the maps being smoothed
on a
Gaussian scale of $4h^{-1}\,\hbox{Mpc}$.

\topinsert
\vbox{\noindent\eightpoint
{Figure 3.}
The mean field ${\bar f}$ (top left panel) and five of the composite
field-constraint
correlation fields $\xi_i({\bf r})$ of a set of constraints (see text). The
panels contain the
contourmaps of the density (top left, contour spacing 0.5) and correlation
values (other 5
panels) in a
$5h^{-1}\,\hbox{Mpc}$ slice through the center of a $100h^{-1}\,\hbox{Mpc}$
box. The spectrum of
the field is the standard CDM spectrum. The 5 field-constraint correlation
functions are a) top
middle: $\langle f\,f_G\rangle$ (contour spacing 0.1), b) top right: $\langle f
\,\nabla_x f_G \rangle$ (contour spacing 0.03) c) bottom left: $\langle f\,
\nabla^2_x
f_G \rangle$ (contour spacing 0.02), d) bottom middle: $\langle
f\,v_{G,x}\rangle$
(contour spacing 0.03) and e) $\langle f\,\sigma_{xx}\rangle$ (contour spacing
0.015), with $f_G$ the value of the smooth density field, $v_{G,x}$ the
$x$-component
of the peculiar velocity and $\sigma_{xx}$ the $xx$-component of the shear
tensor, all
evaluated at the center of the box.
}\endinsert

All the constraints are defined on a Gaussian scale of $4h^{-1}\,\hbox{Mpc}$. A
triaxial peak,
with axis ratio 10~:~9~:~7, of height $f_G=3\sigma_0$ and local density field
curvature
$\nabla^2 f_G=\langle x \rangle \sigma_2 \approx 3.481 \sigma_2$ is positioned
at the center
of the box. Its major axes are slightly oriented with respect to the coordinate
axes of the
box. In addition to these local constraints, there are constraints on the local
gravity and tidal
field. Because we limit ourselves to growing mode linear perturbations we
specify these
constraints in terms of the peculiar velocity and shear. The total
peculiar velocity of the peak is $1145\,\hbox{km/s}$, towards a direction
$26.6^{\circ}$
``north'' of the positive $x$-axis and $22.6^{\circ}$ out of the $x-y$ plane,
in the positive
$z$-direction (note that the specified numerical values of the constraints are
the linear
extrapolations to the present epoch, $a=1$). This corresponds to a value of
$2.00$ times the
velocity dispersion of peaks on a scale of $4h^{-1}\,\hbox{Mpc}$, or $1.66$
times the velocity
dispersion of an average field point on this scale (the lower value of peak
velocities in
comparison with the velocity of field points is due to the extra component
corresponding to
accretion onto the peaks). The shear tensor at the location of the peak is
orientated so that the
off-diagonal terms are zero. The diagonal term in the $x$-direction,
$\sigma_{xx}$, has the
largest magnitude and is positive (dilation), while $\sigma_{yy}$ and
$\sigma_{zz}$ are equal and
negative (contraction). For illustrative purposes we have chosen a rather
extreme value for the
magnitude of the largest element of the shear tensor: $100\,\hbox{km/s/Mpc}$ on
the scale of
$4.0h^{-1}\,\hbox{Mpc}$, $\approx 6.8$ times the dispersion ($\approx
14.5\,\hbox{km/s/Mpc}$) for
the diagonal shear components for peaks (Bond 1987). A good idea of the order
of magnitude of
these shear tensor values is obtained by comparison with the value of the
expansion scalar for
the $3\sigma_0$ peak, $\nabla \cdot {\bf v}_G=-142.47\,\hbox{km/s}$.

The specified constraints can be easily recognized in the resulting mean field,
in the top left
panel of figure~3 (contour spacing 0.5, and the positive value solid contours
separated from the
negative value dotted contours by the thick solid line corresponding to $f=0$).
The contours
around the center of the box clearly reveal the presence of the elongated peak,
oriented with
respect to the coordinate axes. The global density field in the box reflects
the gravity and
tidal field constraints. The source of the motion of the peak is the
concentration of
mass in the upper righthand quarter of the frame, while the clearly discernable
quadrupolar
component in the matter distribution induces the tidal field. To understand how
the different
constraints conspire to produce this mean field it is quite revealing to study
the
individual field-constraint correlation functions $\xi_i({\bf x})$. The five
illustrated
correlation functions $\xi_i({\bf x})$ are the correlation of the field $f({\bf
x})$ with
(1) the value of the smoothed density at the peak position ${\bf x}_i$,
$f_G({\bf x}_i)$
(top middle panel), (2) the value of the first derivative of the smoothed
density field
$\nabla_x f_G ({\bf x}_i)$ (top right panel), (3) the value of the second
derivative
of the smoothed density field $\nabla^2_x  f_G ({\bf x}_i)$ (bottom left
panel),
(4) the peculiar velocity $v_{G,x}({\bf x}_i)$ (bottom middle panel) and (5)
the shear component
$\sigma_{xx}({\bf x}_i)$ (bottom right panel).

The first correlation function (top middle panel) is spherically symmetric,
with a value of 1.0
near the centre, and radially decreasing to a value of 0.0 at the outer contour
(contour spacing is 0.1). Effectively, this correlation function is the
convolution between the
field correlation function $\xi({\bf x})=\langle f f\rangle$ and the Gaussian
filter
function defining the scale of the constrained object. In a similar fashion
we can consider the second correlation function (top right panel, contour
spacing 0.03) to be the
convolution of the correlation function $\xi({\bf x})$ with the first
derivative of the filter
function. This introduces the anisotropy along the $x$-axis, with, within
distances comparable
to the correlation radius, negative values on the left side of the peak and
positive values to
the right. Further outward the correlation function $\xi({\bf x})$ becomes
negative, resulting
in the sign reversal of $\xi_i({\bf x})$. The third correlation function
(bottom
left panel, contour spacing 0.02) is essentially the convolution of the
field correlation function $\xi({\bf x})$ with the second derivative of the
Gaussian
function, $\partial^2 W_G/\partial^2 x$. Because this derivative has two
zero-points along the
$x$-axis we see a negative value near the centre, changing to positive on both
sides of the
centre.

The correlation functions corresponding to the velocity and shear constraints
display
familiar patterns. The function corresponding to the constraint on the peculiar
velocity in the $x$-direction, $v_x$, (middle bottom panel, contour spacing
0.03) is a dipolar
function centred on the position of the peak, with positive values to the
righthand side of the
peak (in the $x$-direction) and negative values to the left. This is evidently
related to the
fact that such a dipolar matter distribution would produce a net gravitational
acceleration, and
corresponding peculiar velocity, in the $x$-direction. In addition, we see that
the constraint
on the shear component $\sigma_{xx}$ results in a clear quadrupolar pattern of
the correlation
function (right bottom panel, contour spacing 0.015). As in the case of the
velocity-field
correlation function, this is related to the non-zero tidal force
$xx$-component that would be
produced by such a quadrupolar mass distribution.

Superposition of the complete set of these correlation fields $\xi_i({\bf x})$,
with the
appropriate weight factors, proportional to the corresponding constraint values
$c_i$,
produces the mean field in the top left panel. Comparison of the different
panels in
figure~4 shows that several of the correlation function patterns can indeed be
recognized
in the mean field.

\topinsert
\vbox{\noindent\eightpoint
{Figure 4.}
The variance of constrained random field realizations. The mean field ${\bar
f}$ and four
different field realizations of a set of constraints (see figure 3) are shown
in the
top left panel and the middle and right rows respectively. The panels contain
the density
contourmaps in the $5h^{-1}\,\hbox{Mpc}$ thick central slice in a
$100h^{-1}\,\hbox{Mpc}$
box. The contour spacing is 0.5. The bottom left panel is the contourmap of the
value of the variance of the field realizations inside the slice, running from
0.0 at
the centre to $\sigma_0\approx 0.95$ at the edge of the box (contour spacing
0.05).
}\endinsert

\medskip
\@{\bf{5.2 Variance of realizations}}
\smallskip
\@After having constructed the mean field ${\bar f}$ (former subsection), we
need to add a
properly sampled residual field realization (eq.~26). Figure~4 provides an idea
of the
possible variations between the residual field realizations and, specifically,
the resulting full
field realizations. In addition to the mean field
illustrated in the the top left panel, four different realizations are shown in
the middle and
right row of panels. All these panels are density contour maps (contour spacing
0.5) in the
same central $5h^{-1}\,\hbox{Mpc}$ thick slice used in figure~3. From the four
field realizations
we can infer that, for example, the mass concentration to the right,
responsible
for the peculiar motion of the peak, can vary substantially in position, shape,
size and
substructure. Moreover, the morphology and distribution of mass clumps inside
the band of matter
along the $x$-axis, main contributor to the specified shear, displays an even
larger variation,
in particular at large distances from the peak.

An analytic expression for the variance of the residual field at any position
${\bf x}$
follows immediately from the independence of the residual field distribution
function from the
numerical values of the imposed constraints (eq.~32), see Appendix~D,

$$\langle F^2({\bf x})|\Gamma\rangle \,=\, \sigma_0^2\,-\,\xi_i({\bf x})
\xi_{ij}^{-1} \xi_j({\bf x}),\eqnew$$

\@with

$$\sigma_0^2=\langle f^2({\bf x}) \rangle,\eqnew$$

\@the variance of the density field (recall that both $f$ and $F$ have zero
mean). The expression in equation~(111) shows that $\langle F^2({\bf x})|\Gamma
\rangle$ is
dependent on ${\bf x}$, and therefore implies the residual field $\langle
F^2({\bf x})|\Gamma
\rangle$ is neither homogeneous nor isotropic. Note that because $F({\bf x})$
is a Gaussian
random field its distribution functional ${\cal P}\left[F|\Gamma\right]$ is
completely specified
by the variance $\langle F^2({\bf x})|\Gamma \rangle$.

The lower left panel shows a contour map of the variance field corresponding to
the
constraint set in the example. Notice the perfect spherical character of this
variance field,
increasing radially outward from the position of the peak, where it is equal to
0.0,
to the general field value $\sigma_0 \approx 0.95$ (contour spacing 0.05). At
first
sight this might seem counterintuitive, as most of the applied constraints are
non-isotropic. However, from equation~(111) we see that $\langle F^2({\bf
x})|\Gamma\rangle$
involves a product of all field-constraint correlation functions $\xi_i({\bf
x})$, independent of
the actual numerical values $c_i$ of the constraints. In our example all 18
peak constraints
have been specified. This means that the anisotropy introduced by e.g. the
dipole distribution
corresponding to the $v_x$ constraint gets fully compensated by the equally
strong $y$ and
$z$ dipole distributions of the $v_y$ and $v_z$ field-constraint correlation
functions. The same
is true for the quadrupole distributions of the shear constraints, as well as
for the correlation functions corresponding to the three first derivatives
$\partial f_G/
\partial x_k$ and the six second derivatives $\partial^2 f_G/\partial x_k
\partial x_l$.

The predicted variance field can also be recognized when comparing the four
field realizations.
They show very small differences in the neighbourhood of the peak, but
further outward the differences become larger and ultimately are equal to the
variations in
any average field.

\medskip
\@{\bf{5.3 Realizations for Gravity and Tidal Field constraints}}
\smallskip
An important ingredient of our code is the ability to put constraints on the
peculiar
gravity or the tidal field acting on a peak. While the peaks in figure~1 do not
have
constraints on either the gravity and tidal field, we intend to give an
impression of
the consequences for both density and velocity fields of imposing such
constraints by
means of a sequence of four random field realiztions. Each of the four examples
contain the same
peak at the center of the box, but differ in the constraints on the gravity and
tidal field to
which the peak is subjected. The central $3\sigma_0$ peak is defined on a
Gaussian scale of
$4h^{-1}\,\hbox{Mpc}$, is spherical in shape, and has a peak curvature of
$\nabla^2 f_G=\langle x \rangle \sigma_2 \approx 2.901 \sigma_2$. By using
the same random number generator for each of the realizations we try to keep
the
differences between the residual fields at a minimum (however, note from
eq.~111 that there will
be differences depending on which constraints are applied).

In the first example (A, figure~5a) the central peak is not subjected to any
velocity and
shear field constraints. In the case of the second example (B, figure~5b), the
same peak is
constrained to have a peculiar velocity of $1000\,\hbox{km/s}$ in the positive
$x$-direction
(note that the specified numerical values of these quantities are the linear
extrapolations to
the present epoch, $a=1$, and that we specify the gravity and tidal field
constraints in
terms of peculiar velocity and shear). In the third realization (C, figure~5c)
we constrain
the shear at the peak's position, while its peculiar velocity is unconstrained.
The
off-diagonal terms of the shear tensor are zero, while $\sigma_{xx}$ has a
positive value
of $50\,\hbox{km/s/Mpc}$ on the scale of $4.0h^{-1}\,\hbox{Mpc}$ and
$\sigma_{yy}$ and
$\sigma_{zz}$ have equal and negative values. In the final, fourth, realization
(D, figure~5d) we
combine the constraints to the peculiar velocity in the second example and the
shear at the
position of the peak in the third example.

The density and velocity field realizations for the four different constraint
sets are the
subject of figure~5. In all four cases we use a set of six panels to highlight
different
aspects of the fields, with each panel illustrating a density or velocity field
in the
same $5h^{-1}\,\hbox{Mpc}$ planar section along the $z$-direction, centered
halfway in the
simulation box. The different contributions to the constrained density fields
are shown
in the top row panels, in combination with the corresponding velocity fields in
the
bottom row. The top left panel contains the contourmap of the mean density
field, smoothed by a
Gaussian filter with  a scale of $2h^{-1}\,\hbox{Mpc}$ (contour spacing
is equal to 0.65=$0.376 \sigma_0(2h^{-1}\,\hbox{Mpc}$)). The corresponding mean
peculiar velocity field is represented by the vector velocity field in the
panel below.
The arrows are the projections of the velocity vectors, for presentation
purposes we limit
outselves to show them at the positions of the gridpoints of a $32^3$ grid.
The length of each arrow is proportional to the magnitude of the velocity, a
length
of $1/20th$ of the boxlength corresponding to a velocity of
$1000\,\hbox{km/s}$.
The corresponding full density field realization is represented by two panels,
a density contour
map of the density field (top middle panel), smoothed on the constraint scale
of
$4h^{-1}\,\hbox{Mpc}$, and a Zel'dovich particle distribution (top right
panel). The constraints
will heavily influence the wavevectors on a scale comparable to and larger than
the scale on
which they are imposed, while the smaller scale waves, responsible for the
subclumps and other
small scale features, are not very much affected due to their negligible
correlation with the
imposed constraints (compare eq.~38 and the listing of constraint kernels
${\hat H}({\bf k})$ in
Appendix~F). The contourmap in the top middle panel (contour spacing
0.275=$0.290\sigma_0(4h^{-1}\,\hbox{Mpc}$)) is therefore the best illustration
of that part of the
density field affected by the constraints. The particle distribution, on the
other hand, shows
the contribution of the small scale waves to the density field at highest
possible resolution.
The particle positions were obtained by using the Zel'dovich approximation to
evolve an initial
distribution of $64^3$ particles to an expansion factor $a=0.4$, approximately
the time at which
the maximum density fluctuation on the scale of 1 gridcell is equal to 10.0. An
additional
advantage of this particle distribution is that it provides a good
representation of how the
density field evolves deep into the quasi-linear regime. The velocity vector
field in the
bottom right panel is the unsmoothed full velocity field realization, and is
closely related
to the Zel'dovich particle distribution. The resolution of this velocity field
representation
is essentially that of one gridcell in the $64^3$ grid that was used to perform
the
constrained field calculations. Filtering this velocity field with a Gaussian
function
of radius $4h^{-1}\,\hbox{Mpc}$ yields the velocity field in the bottom middle
panel,
corresponding to the smoothed density field in the panel above it.

A perfectly spherical density distribution around a maximum at the center of
the box is
evidently the mean density field in example~A, with pure spherical infall
characterizing the
vector velocity field (left row of figure~5a). In a technical sense, recalling
the
discussion on figure~3, we can understand the
spherical density field as the superposition of the spherical correlation
function
$\langle f\, f_G \rangle$ (top middle panel) and three equally large
contributions from
the correlation functions $\langle f\,\nabla^2_x  f_G \rangle$,
$\langle f\,\nabla^2_y  f_G\rangle$, and $\langle f\,
\nabla^2_z f_G\rangle$ (bottom left panel), whose main effect is to produce a
slightly
flatter peak.
The spherically shaped peak can also be recognized in the center of the full
field realization.
However, the shape of the central clump becomes very irregular further outward
from the center.
A comparison with the Zel'dovich particle distribution shows that this clump
consists of at least
four separate subclumps. Note that the central peak, unlike the peak in the
mean field, has a
considerable peculiar motion in the negative $y$ direction, and a small but
nonzero shear. Both
are introduced via the residual field. The absence of correlations between the
small scale waves
is well illustrated by the full velocity field in the bottom right panel of
figure~5a, which
besides spherical infall does not appear to display any additional features but
the expected
noise.


\topinsert
\vbox{\noindent\eightpoint
{\bf Figure 5.} Four different realizations of constrained random fields in the
standard cold
dark matter scenario ($\Omega=1.0, h=0.5$). The constraints are specified on a
Gaussian scale of
$4h^{-1}\,\hbox{Mpc}$. In all cases there is the same $f_G=3\sigma_0$ spherical
peak, with
standard curvature $\nabla^2 f_G \approx 2.901 \sigma_2$, at the center of the
box. In (a) no
further constraints are specified. In (b) the peak is constrained to move with
a peculiar velocity
of $1000\,\hbox{km/s}$ towards the positive $x$-direction. In (c) the diagonal
components
of the traceless shear tensor are constrained to have the value
$\sigma_{xx}=100\,
\hbox{km/s/Mpc}$ and $\sigma_{yy}=\sigma_{zz}=-50\,\hbox{km/s/Mpc}$ while the
off-diagonal
components are all
zero. In (d) the spherical peak has the combined velocity and shear constraints
of (b) and (c).
The four examples are illustrated by six panels. All show an aspect of the
density or
velocity field in the $5h^{-1}\,\hbox{Mpc}$ thick central slice of the
$100h^{-1}\,\hbox{Mpc}$
box. Top left panel: the $2h^{-1}\,\hbox{Mpc}$ smoothed density contourmap of
the mean field
${\bar f}$, contour spacing 0.65. Top middle panel: the $4h^{-1}\,\hbox{Mpc}$
smoothed density
contourmap of the constrained field realization $f$, contour spacing 0.275. Top
right panel:
Zel'dovich particle distribution at the epoch for which the maximum density
fluctuation is
$f=10.0$ on the scale of one gridcell. Bottom left panel: mean velocity vector
map corresponding
to mean density field ${\bar f}$. The vectors are the projected velocity
vectors in this plane.
A vector with a length of $1/20th$ of the boxsize represents a velocity of
$1000\,\hbox{km/s}$.
All velocity vector maps were determined on a $64^3$ grid, but for presentation
purposes only
the vectors on the gridpoints of a $32^3$ subgrid are shown. Bottom middle
panel: vector
map of the constrained velocity field, Gaussian smoothed on a scale of
$4h^{-1}\,\hbox{Mpc}$.
Bottom right panel: unsmoothed constrained velocity field vector map.
}\endinsert

The character of the field realization changes considerably by adding the extra
constraint
that the central spherical peak has a peculiar velocity of $1000\,\hbox{km/s}$
in the
$x$-direction (example~B, figure~5b). The presence of the central spherical
peak can still be
recognized in the mean density field and the full field realization. At the
same time we see that
the global matter distribution is sculpted into the dipolar pattern that
induces the net
gravitational acceleration corresponding to the required peculiar velocity. The
mean velocity
field in the neighbourhood of the central denstiy peak clearly reflects the
required
bulk motion. This local motion is part of a more global pattern in the velocity
field,
consisting of a convergence towards one point, `attractor', in the right half
and a
an outflow pattern from the underdense regions in the left half. Besides this
mean component,
the full velocity field realization contains additional local features, clearly
visible in
the lower middle and right panel of figure~5b. Note that there are several
local regions
from which matter is streaming away, some of these local density depressions
are not
even underdense (note e.g. the saddlepoint around
$[x,y]=[70.0,50.0]h^{-1}\,\hbox{Mpc}$).
Also remark the fact that the central peak is more compact than in example~A,
mainly due to the
very steep density falloff of the peak on the side where it is lying on the
boundary of the
underdense region. This pattern finds its origin in the extra superposition of
the dipolar pattern
characteristic of the correlation function $\langle f \, v_{G,x} \rangle$ (see
figure~3). Equally
striking are the consequences of imposing extra constraints, in example~C, on
the tidal
field and/or corresponding shear at the peak position (figure~5c). The
constraints induce
the expected global quadrupolar mass distribution in the mean density field,
superimposed on the
local spherical peak density distribution. The band of matter parallel to the
$x$-axis,
visible in both the mean and final density field, induces the dilational
shearing motion along
the $x$-direction and the compressional shear along the other two directions,
in collaboration
with the underdense regions below and on top of it. The presence at the peak
position of the
positive $\sigma_{xx}$ component, along with the negative $\sigma_{yy}$
component of half its
magnitude, is most strikingly visible in the mean velocity field. In the full
field realization
we can also recognize the presence of other components than the quadrupolar
one. The central
high-density ridge is littered with numerous small scale peaks of different
sizes (see e.g
the Zel'dovich particle distribution) while a clear dipolar component can also
be discerned
in the density distribution. High-density regions are concentrated in the lower
half of the
box, inducing the sizable peculiar motion of the peak towards the negative
$y$-direction
that can be seen in the velocity field realizations in the lower middle and
right panels.
Finally, figure~5d shows how the combination of the constraints on the peculiar
velocity and the
shear in example~D work out. The corresponding mean density field clearly
contains both a dipolar
and a quadrupolar component, both of which are also conspicuously present in
the full density
field realization (also compare with the Zel'dovich particle distribution). In
both the mean
velocity field and the full velocity field realization we can recognize the
specified
peculiar velocity and shear at the position of the central peak. The particle
distribution
shows that the clumps on the right hand side of the center are more massive
than the ones in
figure~5c. The agglomerate of these clumps conspires to form a big attractor,
easily
recognizable, that induces the large peculiar motion of the peak.

\medskip
\@{\bf {\bigone 6. Summary and Discussion}}
\smallskip
\@In this paper we have developed a formalism to set up cosmological initial
Gaussian
random density and velocity fields that can contain one or more peaks or dips,
with the
intention to generate appropriate initial conditions for cosmological $N$-body
simulations
that focus on the evolution of the progenitors of the present-day galaxies and
clusters and
their environment. The method is suited for fields with any arbitrary power
spectrum $P(k)$.
Central objective of our algorithm is the ability to sculpt the local and
global matter
distribution in a sufficiently large volume such that certain physical
characteristics of the
density and velocity field in the immediate neighbourhood of the primordial
peaks have a priori
specified values. The generation of these constrained density fields is an
application and
elaboration of the the Hoffman \& Ribak (1991) prescription. They showed that
there is a simple
and elegant solution to achieve this if the constraints are linear functionals
of the field.
We have presented the implementation of our method following a comprehensive
discussion of the
fundamentals underlying their method.

A maximum of 21 characteristics is used to specify the density and velocity at
and around the
position of the peak. They can be divided into three groups:

[1] The scale and position of the peak. We identify a peak as a local maximum
in the density field
that has been smoothed by a Gaussian filter function with a characteristic
scale $R_G$,
although the formalism can be very easily extended to other filter functions.
The peak may
be positioned at an arbitrary position within the simulation box.

[2] The local density field. In total 10 constraints are needed to fully
specify the
density field in the immediate vicinity of the peak. The first one concerns the
height of the
peak. In addition, three constraints are needed to assure that the three first
derivatives
of the smooth density field vanish at is summit. Finally, the six second order
derivatives of the smooth density field are set by specifying the compactness
$\nabla^2 f_G$, the axis ratios and the orientation of the peak.

[3] The local gravitational field. The specification of the gravitational field
around the peak
introduces 8 additional constraints: the three components of the smoothed
peculiar
gravity at the location of the peak and the five independent components of the
traceless tidal tensor. The resulting density field is sculpted in such a way
that it
induces the desired amount of net gravitational and tidal forces. We usually
restrict ourselves to the growing mode component of the density field. In the
linear
clustering regime the peculiar gravity and tidal field are therefore directly
proportional
to the peculiar velocity and the shear, so that we commonly use the latter to
specify the
gravitational field constraints.

It may be worthwhile to point out that in a linear density fluctuation
field several of the above quantities are correlated. For example, we
find that there is a strong correlation between the tidal field
tensor and the mass tensor, expressing itself in the tendency of the
tidal field to align itself along the principal axes of the mass
tensor.

The constraints that we consider here are linear functionals of the density
fluctuation
field $f$, and therefore can be written as convolutions of the field with a
specific
function. Consequently, it is most convenient to perform the relevant
calculations in
Fourier space. The generation of a constrained field realization basically
consists of
the sum of an arbitrary field realization with the convolution of the power
spectrum with
a function that is the weighted sum of the different constraint kernels, the
weights depending
on the specified values of the constraints and the values of the constraints
for the
unconstrained field (see eq.~26). The expressions for these constraint kernels
are
derived from the particular constraints to which they are related. In
Appendix~F we list the
kernels used in our code.

The Hoffman-Ribak algorithm that we have described here is considerably faster
and more
generally applicable than the original Bertschinger (1987) algorithm. Its
superior speed
is due to the direct and simple way of sampling the residual field, rendering
an
iterative ``simulated annealing'' technique superfluous. Moreover, because it
is a direct
method it has the additional advantage of superior accuracy. Extensive testing
of constrained field realizations showed that the implementation is very
precise, leading
to accuracies in the order of $0.01\%$ for the imposed quantities.
In the computer implementation of our code the constrained field is evaluated
on a
periodic three-dimensional lattice. This has the advantage of being able to
perform
the Fourier transforms by means of a Fast Fourier Transform, with the advantage
of
being considerably faster than methods based on a direct Fourier transform. A
disadvantage
of the FFT is that they have a rather weak sampling at low $k$, while direct
Fourier
transforms enable a far better sampling in that range. In their multipole
constrained field
method Bond and Meyers (1993) therefore resort to direct Fourier transforms,
resulting
in an excellent sampling at low and intermediate $k$.

In addition to the fact that the Hoffman-Ribak method provides us with a fast,
efficient and
accurate method to generate constrained random fields it has two other
important advantages.
The first one is that the implementation of a large variety of constraints is
relatively
straightforward through the convolution integrals in Fourier space. Secondly,
unlike most
other efficient algorithms it is equally suitable and efficient for local and
non-local constraints. Although the illustrations of the peak constraints in
section~5
were mainly local in character, centered on one peak, the developed formalism
allows the
generation of numerous peaks and dips at different positions (see figure~1).

In our application to peaks we followed the philosophy that each of the
constraints
corresponds to a different physical quantity. Another class of possible
applications of
the Hoffman-Ribak procedure is the reconstruction of (linear) density fields
from the
measurement of the same physical quantity at several different positions inside
a certain
volume. A nice illustration of this is the work by Ganon \& Hoffman (1993), who
reconstructed the the density field in the ``local'' universe from the observed
velocity field
sampled at 181 different positions within a sphere of $40h^{-1}\,\hbox{Mpc}$
around us, assuming
that it is a realization of a standard cold dark matter field. They showed that
the method
recovers the main features of POTENT's density field (Dekel, Bertschinger \&
Faber 1990),
in particular the Great Attractor region. The interesting feature of this
reconstruction
application is that it creates high-resolution fields subject to the
low-resolution data,
for the given underlying model. It therefore offers the charming and
interesting opportunity to set up initial conditions for $N$-body simulations
from observations
of the local Universe, so that the nonlinear evolution of our ``local''
Universe in a
particular cosmological scenario can be studied. A related and promising
application would be the
construction of high-resolution microwave background maps from the the
large-scale anisotropies
measured by COBE (Bunn et al. 1994).

This class of constraint problems, where the constraints consist of the value
of the same
physical quantity $\psi({\bf r})$ at many different positions, offers the
advantage that for
every constraint the constraint-field correlation function $\xi_i({\bf
r})=\langle
\psi({\bf r}_i) f({\bf r})\rangle \equiv \Upsilon({\bf r}-{\bf r}_i)$ (see
eq.~24) can be
evaluated from the same general correlation function $\Upsilon({\bf x})$. The
same is true for
the constraint-contraint correlation function $\xi_{ij}$. In particular, this
will be a great
advantage if the constraint values are imposed at equally spaced points on a
grid. This is
the approach followed by Ganon \& Hoffman (1993), who determined the constraint
values
for the velocity potential on a grid by spatial interpolation from observed
values of the
peculiar velocity. The computation of the required values of $\xi_i({\bf x})$
and the inverse
constraint-constraint correlation matrix $\xi_{ij}^{-1}$ can then be simply
accomplished by
two FFTs. This can be easily seen from the following. Because the quantity
$\psi({\bf x})$ is a linear functional of the density field $f({\bf x})$, its
Fourier transform
${\hat \psi}({\bf k})$ is a product of the Fourier transform ${\hat f}({\bf
k})$ of the field
$f({\bf x})$ with a kernel function ${\hat h}({\bf k})$, ${\hat \psi}({\bf
k})={\hat h}({\bf k})
{\hat f}({\bf k})$. Examples of such fields $\psi({\bf x})$ are the
gravitational potential, the
peculiar velocity in the linear regime, or the temperature variations in the
cosmic background
radiation field. After evaluating the corresponding expressions for ${\hat
h}({\bf k})$ at
wavenumbers ${\bf k}_p$ (compare the kernel functions listed in Appendix~F),
the values of
$\xi_i({\bf x}_j)=\langle \psi({\bf x}_i) f({\bf x}_j)\rangle$ and
$\xi_{ij}=\langle \psi({\bf x}_i) \psi({\bf x}_j) \rangle$ can be found from
$\xi_i({\bf x}_j)=\Upsilon({\bf x}_i-{\bf x}_j)$ and $\xi_{ij}=\Psi({\bf
x}_i-{\bf x}_j)$,
where

$$\eqalign{\Upsilon({\bf x})&={\displaystyle 1 \over \displaystyle N}
\sum_{p=0}^{N-1}\,
{\hat h}({\bf k}_p) P(k_p)\,e^{-i{\bf k}_p\cdot{\bf x}}\cr
\Psi({\bf x})&={\displaystyle 1 \over \displaystyle N} \sum_{p=0}^{N-1}\,
{\hat h}({\bf k}_p)^2 P(k_p)\,e^{-i{\bf k}_p\cdot{\bf x}}\cr}\,.\eqnew$$

\@In fact, the inverse matrix of $\xi_{ij}$ can be found directly and very
simply from
$\xi^{-1}_{kl}=\Theta({\bf x}_k-{\bf x}_l)$, where $\Theta({\bf x})$ is the
inverse
of $\Psi({\bf x})$, i.e. $\Psi({\bf x}_k-{\bf x}_i) \Theta({\bf x}_i-{\bf x}_l)
= \delta_{kl}$,
and therefore given by the Fourier sum

$$\Theta({\bf x})={\displaystyle 1 \over \displaystyle N}
\sum_{p=0}^{N-1}\,{\displaystyle 1
\over \displaystyle {\hat h}({\bf k}_p)^2 P(k_p)}\,e^{-i{\bf
k}_p\cdot{\bf x}}\,.\eqnew$$

\@The computation of the discrete Fourier sums $\Upsilon({\bf x})$ and
$\Theta({\bf x})$ is
accomplished by a FFT, so that the computational cost is only ${\cal O}(N \log
N)$.
Note that because of the periodic boundary conditions intrinsic to the FFT each
coordinate
of ${\bf x}_i$ can only attain half of the values along each axis, so that in
total only
${1 \over 8}$ of the computational box is used for the field reconstruction.
Finally,
the independent Fourier components of the unconstrained field ${\tilde f}({\bf
x})$
are generated. The subsequent computation of ${\tilde f}({\bf x})$ itself
demands one FFT, and
the computation of the corresponding constraint values $\tilde c_j$ involves
another FFT (compare eq.~38). Combining all these results in the final
evaluation of
the constrained field according to equation~(35) consists of the computation of
the
double product $\xi_i({\bf x}) \xi^{-1}_{ij} (c_j - {\tilde c}_j)$ for every
point ${\bf x}_j$,
making it an ${\cal O}(N^3)$ formalism. However, unlike the formalism developed
in
section 3, this procedure does not involve a very costly matrix inversion of
$\xi_{ij}$,
implying it to be far more efficient and the method of choice for this
particular class of
applications. On the other hand, when each of the $M$ constraint quantities
have a different
character, concern different scales, arbitrary non-grid positions, or different
filters, this
procedure cannot be straightforwardly applied. In those cases a formalism
similar to the one
presented in this paper is automatically implied.

As a final note we should issue a cautionary remark on the practical
implementation of
our constrained random field code. The initial density fields are set up in a
box with periodic
boundary conditions. This means that the mean density of the box is exactly
equal to the mean density of the Universe. The structure generated within the
box
is therefore not entirely typical, since overdense regions must necessarily be
surrounded by
low-density regions. This need not be true in general, from the theory of
Gaussian random
field we know that peaks tend to cluster. The simulation box should therefore
not be taken
too small, the resulting structure might be very atypical. Evidently, this
conflicts with the
demand to make the box as small as possible to achieve the highest possible
resolution. The
chosen box size should therefore be a compromise between these two.

In summary, we can conclude that the Hoffman-Ribak method provides a powerful
and elegant tool to
study the formation and evolution of specific cosmological objects in great
detail under ideal
conditions. The tools developed in this paper should essentially be regarded to
constitute a
laboratory equipment set\footnote*{our FORTRAN computer code will be
available upon request, and will be incorporated as part of the
COSMICS package, Bertschinger 1995}.
They allow us to set up very specific conditions for the objects under study. A
sequence of
experiments based on a range of different circumstances will subsequently yield
 a maximum of
insight into the systematic dependence of structure formation on specific
physical quantities. By concentrating on one specific application, peaks in the
density field,
we hope to have provided a recipy for constructing similar applications and
extensions
for different quantities in fields of a possibly different character. A
straightforward
extension of our formalism will for example be to consider peaks in the
gravitational
potential field instead of in the density field.

\medskip
\@{\bf {\bigone Acknowledgements.}}
RvdW is very grateful to Bernard Jones for the encouragement
and the many useful instructive conversations and discussions on this
subject. He also wishes to thank Michiel van Haarlem and Arif Babul for the
collaborations and discussions that formed an important impetus to this
project.
In addition, we are indebted to Dick Bond for insightful and encouraging
suggestions. EB acknowledges support from NSF grant AST-9318185 and NASA grant
NAG5-2816.

\bigskip
\@{\bf {\bigone References}}
\parskip=0pt plus1pt
\medskip
\hang Bardeen, J.M., Bond, J.R., Kaiser, N., Szalay, A.S., 1986, ApJ, 304, 15
\hang Bertschinger, E., 1987, ApJ, 323, L103
\hang Bertschinger, E., 1992, in New Insights into the Universe, ed.
V. J. Martinez, M. Portilla \& D. Saez (New York: Springer-Verlag), p. 65
\hang Bertschinger, E., Jain, B., 1994, ApJ, 431, 486
\hang Bertschinger, E., 1995, MIT report-no GC3-033 (astro-ph/9506070)
\hang Binney, J., Quinn, T., 1991, MNRAS, 249, 678
\hang Bond, J.R., Meyers, S.T., 1993, CITA preprint
\hang Bunn, E.F., Fisher, K.B., Hoffman, Y., Lahav, O., Silk, O., Zaroubi, S.,
1994, ApJ,
submitted
\hang Doroshkevich, A.G., 1970, Afz, 6, 320
\hang Feynmann, R., Hibbs, A., 1965, Quantum Mechanics and Path Integrals (New
York:
McGraw-Hill)
\hang Ganon, G., Hoffman, Y., 1993, ApJ, 415, L5
\hang Goldstein, H., 1980, Classical Mechanics (2nd ed., Addison-Wesley)
\hang Hoffman, Y., Ribak, E., 1991, ApJ, 380, L5
\hang Katz, N., Quinn, T., Gelb, J.M., 1993, MNRAS, 265, 689
\hang Peacock, J.A., Heavens, A.F., 1985, MNRAS, 217, 805
\hang Peebles, P.J.E., 1980, The Large-Scale Structure of the Universe
(Princeton
University Press)
\hang Van de Weygaert, R., 1991, Voids and the geometry of Large Scale
Structure,
Ph.D. Thesis Leiden University
\hang Van de Weygaert, R., Van Kampen, E., 1993, MNRAS, 263, 481
\hang Van de Weygaert, R., Babul, A., 1994, ApJ, 425, L59
\hang Van de Weygaert, R., Babul, A., 1995, ApJ, submitted
\hang Van Haarlem, M., Van de Weygaert, R., 1993, ApJ, 418, 544
\hang Zel'dovich, Ya.B., 1970, A\&A, 5, 84

\bigskip
\@{\bf Appendix A.\ \ The intersection of a sphere and a polygon}
\smallskip
\@In section 2.2, equation~19, we saw that imposing the set of constraints
$\Gamma=\{C_i[f;{\bf x}_i]=c_i;\  i=1,\ldots,M\}$ is equivalent to
a change of the action $S[f]$ into

$$2S[f]=\int \int \,f({\bf x}_1)K({\bf x}_1-{\bf x}_2)
f({\bf x}_2)\,\,d{\bf x}_1\,d{\bf x}_2\,-\,C^t \xi_{ij}^{-1} C\,,\eqno\rm(A1)$$

\@with $\xi_{ij}$ the $(ij)^{th}$ element of the matrix ${\bf Q}=\langle C^t
C\rangle$.

The $i^{th}$ constraint $C_i({\bf x}_i)$ (in this appendix we will use the
simplifying notation $C_i({\bf x})$ for $C_i[f;{\bf x}]\,$) can be
written as a convolution with a Dirac delta function $\delta_D({\bf x})$,

$$C_i({\bf x}_i)=\int d{\bf x}_2 \,\delta_D({\bf x}_2)\,
C_i({\bf x}_i-{\bf x}_2)=\int d{\bf x}_1 \int d{\bf x}_2 \, \xi({\bf x}_1)
K({\bf x}_1-{\bf x}_2) C_i({\bf x}_i-{\bf x}_2)\,,\eqno\rm(A2)$$

\@where we have used the fact that $K({\bf x})$ is the functional inverse of
the correlation function $\xi({\bf x})$ (eq.~7, section 2.1). By using the
convolution
theorem we can express this double convolution integral in Fourier space as

$$C_i({\bf x}_i)=\int \d3k\,{\hat C_i}({\bf k}) P({\bf k}) {\hat K}({\bf k})\,
e^{-i {\bf k}\cdot{\bf x}_i}\,,\eqno\rm(A3)$$

\@where ${\hat C}_i({\bf k})$ is the Fourier transform of $C_i({\bf x})$,

$$C_i({\bf x})=\int \d3k
\,{\hat C}_i({\bf k})\,e^{-i{\bf k}\cdot{\bf x}}\,,\eqno\rm(A4)$$

\@and $P({\bf k})=P(k)$, the spectral density, and ${\hat K}({\bf k})$ the
Fourier
transforms of $\xi({\bf x})$ and $K({\bf x})$ respectively
(eq.~B5). The formal definition for the spectral density $P(k)$ is
(Bertschinger 1992)

$$(2\pi)^3 P(k_1)\,\delta_D({\bf k}_1-{\bf k}_2) = \langle {\hat f}({\bf k}_1)
{\hat f}^\ast({\bf k}_2)\rangle \,,\eqno\rm(A5)$$

\@with $\delta_D({\bf k}_1-{\bf k}_2)$ the Dirac delta function. In an
analogous
fashion a function ${\hat P}_i({\bf k})$ can be introduced,

$$(2\pi)^3 {\hat P_i}({\bf k}_1)\,\delta_D({\bf k}_1-{\bf k}_2) = \langle
{\hat C}_i({\bf k}_1) {\hat f}^\ast({\bf k}_2)\rangle \,,\eqno\rm(A6)$$

\@from which we obtain, in combination with equation~(A5), a relation
between ${\hat f}({\bf k})$ and ${\hat C}_i({\bf k})$,

$${\displaystyle \langle {\hat f}({\bf k}_1) {\hat f}^\ast({\bf k}_2)\rangle
\over
\displaystyle P(k_1)} = {\displaystyle \langle {\hat C}_i({\bf k}_1)
{\hat f}^\ast({\bf k}_2)\rangle \over {\hat P}_i({\bf k}_1)} \qquad \Rightarrow
\qquad {\hat C}_i({\bf k})= {\displaystyle {\hat P}_i({\bf k}) \over
\displaystyle P(k)}
\,{\hat f}({\bf k})\,.\eqno\rm(A7)$$

\@By subsequently inserting this relation in the Fourier integral of
equation~(A3),
and using definition~(A6), we get

$$\eqalign{C_i({\bf x}_i)&=\int \d3k\,{\hat f}({\bf k}) {\hat P}_i({\bf k})
{\hat K}({\bf k})\,e^{-i {\bf k}\cdot{\bf x}_i}\cr
&=\int \int {\displaystyle d{\bf k}_1 \over \displaystyle (2\pi)^3}
{\displaystyle d{\bf k}_2 \over \displaystyle (2\pi)^3}\,{\hat f}({\bf k}_1)
{\hat K}({\bf k}_1) \langle {\hat C}_i({\bf k}_1) {\hat f}^\ast({\bf
k}_2)\rangle
\,e^{-i {\bf k}_1\cdot{\bf x}_i}\cr
&= \int \int \,f({\bf x}_1)K({\bf x}_1-{\bf x}_2) \xi_i({\bf x}_2)\,d{\bf x}_1
\,d{\bf x}_2\cr}\,,\eqno\rm(A8)$$

\@where the function $\xi_i({\bf x})$ is the field-constraint correlation
function and
the Fourier transform of ${\hat P}_i({\bf k})$,

$$\xi_i({\bf x}) \equiv \langle f({\bf x}) C_i({\bf x}_i)\rangle =
\int \d3k\, {\hat P}_i({\bf k}) e^{-i{\bf k}\cdot({\bf x}_i-{\bf
x})}\,.\eqno\rm(A9)$$

\@Since the field $f({\bf x})$ also obeys the constraints $C_j=c_j$ the
expression
$C_i({\bf x}_i)\, \xi_{ij}^{-1} C_j({\bf x}_j)$ in equation~(A1) can be
replaced by

$$C_i \,\xi_{ij}^{-1} C_j =\int \int\,f({\bf x}_1) K({\bf x}_1-{\bf x}_2) {\bar
f}
({\bf x}_2)\,d{\bf x}_1\,d{\bf x}_2\,,\eqno\rm(A10)$$

\@where we have defined the field ${\bar f}({\bf x})$ by

$${\bar f}({\bf x})\equiv \xi_i({\bf x}) \,\xi_{ij}^{-1} c_j\,.\eqno\rm(A11)$$

\@The constrained action $S[f]$ in equation~(A1) can therefore be written as

$$\eqalign{2S[f]&=\int \int \,f({\bf x}_1)
K({\bf x}_1-{\bf x}_2)\Bigl\{f({\bf x}_2)-{\bar f}({\bf x}_2)\Bigr\}
\,d{\bf x}_1 d{\bf x}_2\cr
&=\int \int \,\Bigl\{f({\bf x}_1)-{\bar f}({\bf x}_1)
\Bigr\}K({\bf x}_1-{\bf x}_2)\Bigl\{f({\bf x}_2)-{\bar f}({\bf x}_2)\Bigr\}
\,d{\bf x}_1 d{\bf x}_2\ + \cr
&\ \ \,\int \int \,{\bar f}({\bf x}_1)K({\bf x}_1-
{\bf x}_2)\Bigl\{f({\bf x}_2)-{\bar f}({\bf x}_2)\Bigr\}\,d{\bf x}_1 d{\bf
x}_2\,.\cr}
\eqno\rm(A12)$$

\@It can be easily shown that the second term on the right hand side of
equation~(A12)
is equal to zero because

$$\int \int \,{\bar f}({\bf x}_1) K({\bf x}_1-
{\bf x}_2) {\bar f}({\bf x}_2)\,d{\bf x}_1 d{\bf x}_2=
\int \int \,f({\bf x}_1) K({\bf x}_1-{\bf x}_2)
{\bar f}({\bf x}_2) \,d{\bf x}_1 d{\bf x}_2=  c_i\,\xi_{ij}^{-1} c_j
\,.\eqno\rm(A13)$$

\@This relation follows directly from equation~(A10) for the second integral,
while for the
first integral it follows from the fact that

$$\eqalign{\int \int \,{\bar f}({\bf x}_1) K({\bf x}_1-{\bf x}_2)
{\bar f}({\bf x}_2) \,d{\bf x}_1 d{\bf x}_2&= \xi_{ij}^{-1} \xi_{kl}^{-1} c_j
c_l
\,\int \int \,\xi_i({\bf x}_1) K({\bf x}_1-{\bf x}_2) \xi_k({\bf x}_2)\,d{\bf
x}_1
d{\bf x}_2\cr
&=\xi_{ij}^{-1} \xi_{kl}^{-1} \xi_{ki} c_j c_l = c_l\,\xi_{lj}^{-1} c_j \cr}\,.
\eqno\rm(A14)$$

\@where we have used the substitution of the Fourier expression of
$\xi_i({\bf x})$ (eq.~A9) to evaluate the integral,

$$\eqalign{&\int \int \,\xi_i({\bf x}_1) K({\bf x}_1-{\bf x}_2) \xi_k({\bf
x}_2)
\,d{\bf x}_1 d{\bf x}_2=\cr
&\qquad\qquad=\int \int {\displaystyle d{\bf k}_1 \over \displaystyle (2\pi)^3}
{\displaystyle d{\bf k}_2 \over \displaystyle (2\pi)^3}\,{\hat P}_i({\bf k}_1)
{\hat K}({\bf k}_2) {\hat P}_k^{\ast}({\bf k}_2)\,(2\pi)^3 \delta_D({\bf
k}_2-{\bf k}_1)
\,e^{-i{\bf k}_1 \cdot {\bf x}_i} e^{i {\bf k}_2 \cdot {\bf x}_k}\cr
&\qquad\qquad=\int \int {\displaystyle d{\bf k}_1 \over \displaystyle (2\pi)^3}
{\displaystyle d{\bf k}_2 \over \displaystyle (2\pi)^3}\,\langle
{\hat C}_k^{\ast}({\bf k}_2) {\hat C}_i({\bf k}_1)\rangle\,e^{-i{\bf k}_1 \cdot
{\bf x}_i} e^{i {\bf k}_2 \cdot {\bf x}_k}
\ \ =\ \ \langle C_k({\bf x}_k) C_i({\bf x}_i)\rangle \ \ = \ \ \xi_{ki}\,.\cr}
\eqno\rm(A15)$$

\@The transition from the $2^{\rm nd}$ to $3^{\rm rd}$ line in equation~(A15)
has been
made by combining equation~(A6) and (A7),

$$(2\pi)^3 {\hat P}_i({\bf k}_1) \delta_D({\bf k}_1 - {\bf k}_2) =
{\displaystyle P({\bf k}_2) \over \displaystyle {\hat P}_k^{\ast}({\bf k}_2)}\,
\langle {\hat C}_k^{\ast}({\bf k}_2) {\hat C}_i({\bf k}_1)
\rangle\,,\eqno\rm(A16)$$

\@and the fact that $P(k)=1/{\hat K}({\bf k})$ (see eq.~B6, app. B).

\vskip 0.2truecm
By defining the ``residual field'' $F({\bf x})=f({\bf x})-
{\bar f}({\bf x})$ we can therefore conclude from equation~(A12) that the
constrained
action $S[f]=S[F]$ can be written as

$$2S[F]=\int d{\bf x}_1 \int d{\bf x}_2\,F({\bf x}_1) K({\bf x}_1 -{\bf x}_2)
F({\bf x}_2)\,,\eqno\rm(A17)$$

\@which is the expression needed in Section 2.2.

\bigskip
\@{\bf Appendix B:  Diagonalisation of the action S[F]}
\smallskip
\@In this appendix we will rewrite the action $S[F]$ (eq. 20),

$$S[F]={\displaystyle {1 \over 2}} \int d{\bf x}_1 \int d{\bf x}_2 F^{\ast}
({\bf x}_1) K({\bf x}_1-{\bf x}_2) F({\bf x}_2).\eqno\rm(B1)$$

\@in terms of the Fourier transform ${\hat F}({\bf k})$ of the fluctuation
field $F({\bf x})$,

$$F({\bf x})=\int \d3k\,{\hat F}({\bf k})
\,e^{-i{\bf k}\cdot{\bf x}}.\eqno\rm(B2)$$

\@The kernel $K({\bf x})$ in equation (B1) is the functional inverse of the
correlation function $\xi(x)$ (eq. 7),

$$\int d{\bf x}\,K({\bf x}_1-{\bf x}) \xi({\bf x}-{\bf x}_2)=
\delta_D({\bf x}_1-{\bf x}_2)\,.\eqno\rm(B3)$$

\@By virtue of the convolution theorem this equation is equivalent to

$$\int \d3k\,{\hat K}({\bf k})
P(k)e^{i{\bf k}\cdot({\bf x}_1-{\bf x}_2)}=\delta_D({\bf x}_1-{\bf x}_2),
\eqno\rm(B4)$$

\@where ${\hat K}({\bf k})$ and $P(k)$ are the Fourier transform of
$K({\bf x})$ and $\xi({\bf x})$ ,

$$K({\bf x})=\int \d3k
\,{\hat K}({\bf k})e^{-i{\bf k}\cdot{\bf x}}\,,\qquad
\xi({\bf x})=\int \d3k
\,P(k) e^{-i{\bf k}\cdot{\bf x}}\,.\eqno\rm(B5)$$

\@The identification of the left part of equation~(B4) with the Fourier
integral expression of the Dirac delta function implies that
${\hat K}({\bf k}) ={1 / P(k)}\,$.
Consequently,

$$K({\bf x})=\int \d3k\,
{\displaystyle 1 \over \displaystyle P(k)}\,e^{-i{\bf k}\cdot{\bf
x}}\,.\eqno\rm(B6)$$

\@Likewise, the insertion of ${\hat K}({\bf k}) ={1 / P(k)}\,$ into
the double convolution of equation~(B1),

$$2S[F]=\int \d3k
\,{\hat F}^\ast({\bf k}) {\hat K}({\bf k}) {\hat F}({\bf k})\,,
\eqno\rm(B7)$$

\@yields the Fourier expression for $S[F]$,

$$S[f]=\int \d3k
\,{\displaystyle |{\bf {\hat F}}({\bf k})|^2 \over \displaystyle 2 P(k)}
\,,\eqno\rm(B8)$$

\@which is equation 29 in section 3.

\bigskip
\@{\bf Appendix C: A heuristic proof that ${\cal P}\left[F|
\Gamma\right]$ is independent of $c_i$.}
\smallskip
\@A field $f({\bf x})$ can be viewed as an $N$-dimensional vector
$(f_1,\ldots,
f_N)$ in $N$-dimensional ``field'' space, with $N \rightarrow \infty$.
The fields $f({\bf x})$ that obey the set of $M$
constraints $\Gamma=\{C_i[f;{\bf x}_i]=c_i;\ i=1,\ldots,M\}$ define an
$(N-M)$-dimensional hypersurface in this $N$-dimensional space. For reasons
of convenience this hypersurface will also be denoted as $\Gamma$.
The only restriction that we impose on the constraints $C_i$ is that
they are linear,

$$C_i[f_1+f_2;{\bf x}]=C_i[f_1;{\bf x}]+C_i[f_2;{\bf x}]\,.
\eqno\rm(C1)$$

\@Each of the hypersurfaces $\Gamma$ contain a special point
${\bar f}({\bf x})$, the mean of the fields satisfying the constraints
$\Gamma$,

$${\bar f}({\bf x})=\langle f({\bf x})|\Gamma\rangle=
\xi_i({\bf x})\,\xi_{ij}^{-1} c_j\,,\eqno\rm(C2)$$

\@where $\xi_i({\bf x})$ is the cross-correlation between the field and the
$i^{th}$ constraint $C_i[f;{\bf x}]$, and $\xi_{ij}$ the correlation
between the $i^{th}$ and $j^{th}$ constraints, $C_i[f]$ and $C_j[f]$ (notice
that in this notation we stress the functional character of the constraints).
Both $\xi_{ij}$ and $\xi_i({\bf x})$ are defined in equation~(24)
(Section 2.2). Each of the fields $f({\bf x})$ in $\Gamma$ have a
corresponding residual field $F({\bf x})$, defined as the difference
between the field $f({\bf x})$ and the mean field ${\bar f}({\bf x})$ of
$\Gamma$,

$$F({\bf x})\equiv f({\bf x})-{\bar f}({\bf x})\,.\eqno\rm(C3)$$

\@Imagine two arbitrarily chosen constraint hypersurfaces, the
first one corresponding to the constraint set $\Gamma_1=\{C_i[f;{\bf
x}_i]=c_{i,1};
\ i=1,\ldots,M\}$ and the other one to the set $\Gamma_2=\{C_i[f;{\bf x}_i]=
c_{i,2};\ i=1,\ldots,M\}$. The mean fields of the sets $\Gamma_1$ and
$\Gamma_2$ are ${\bar f}_1$ and ${\bar f}_2$.
Consider the translation of an arbitrary field $f_1({\bf x}) \in
\Gamma_1$ by a field $T_{2,1}({\bf x})$ into a field $f_T({\bf x})$,

$$f_{T}({\bf x}) \equiv f_1({\bf x})+T_{2,1}({\bf x})\,,\eqno\rm(C4)$$

\@where the translation $T_{2,1}({\bf x})$ is defined by

$$T_{2,1}({\bf x})\equiv {\bar f}_2({\bf x})-{\bar f}_1({\bf x})=
\xi_i({\bf x})\,\xi_{ij}^{-1} (c_{j,2}-c_{j,1})\,.\eqno\rm(C5)$$

\@This definition of $T_{2,1}$ immediately implies that the mean field
${\bar f}_1({\bf x})$ of $\Gamma_1$ is transformed into the mean field
${\bar f}_2({\bf x})$ of $\Gamma_2$. From equations~(C4) and (C5)
and the linearity of the constraints $C_i$ we can infer that

$$\eqalign{C_i[f_{T}]&=C_i[f_1]+C_i[T_{2,1}]\cr
&=C_i[f_1]+C_i[{\bar f}_2]-C_i[{\bar f}_1]\cr
&=c_{i,1}+c_{i,2}-c_{i,1}\ =\  c_{i,2}\,,\cr}\eqno\rm(C6)$$

\@The field $f_T({\bf x})$ therefore obeys the constraint set $\Gamma_2$.
This is true regardless of the field $f_1({\bf x}) \in \Gamma_1$. Moreover,
the inverse translation $-T_{2,1}$ transforms the resulting field
$f_2({\bf x})$ back into $f_1({\bf x})$. The two hypersurfaces $\Gamma_1$
and $\Gamma_2$ are therefore linked by a one-to-one mapping, so that

$${\cal P}[f_1|\Gamma_1]={\cal P}[f_2|\Gamma_2]\,,\eqno\rm(C7)$$

\@where ${\cal P}[f_1|\Gamma_1]$ is the probability of having a
specific field $f_1({\bf x})$ under the condition that they satisfy the
constraints $\Gamma_1$, and $f_2({\bf x})$ is the field in the
hypersurface $\Gamma_2$ that is linked to $f_1({\bf x})$ by the
translation $T_{2,1}({\bf x})$ (eq.~C4). The conditional probabilities
for the corresponding residual fields $F_1({\bf x}) \equiv (f_1({\bf x})-
{\bar f}_1({\bf x}))$ and $F_2({\bf x}) \equiv (f_2({\bf x})-
{\bar f}_2({\bf x}))$ can be inferred from equation~(C7),

$${\cal P}[F_1|\Gamma_1]={\cal P}[f_1|\Gamma_1]={\cal P}[f_2|\Gamma_2]=
{\cal P}[F_2|\Gamma_2]\,.\eqno\rm(C8)$$

\@Finally, consider the transformation of the residual field
$F_1({\bf x})$ under the translation $T_{2,1}$,

$$\eqalign{F_1({\bf x})&\equiv f_1({\bf x})-{\bar f}_1({\bf x})\cr
&=(f_1({\bf x})+T_{2,1}({\bf x}))-({\bar f}_1({\bf x})+T_{2,1}({\bf x}))\cr
&=f_2({\bf x})-{\bar f}_2({\bf x})\ =\ F_2({\bf x})\,.\cr}
\eqno\rm(C9)$$

\@In other words, the residual field $F({\bf x})$ is invariant under the
translation $T_{2,1}$, i.e. $F_1=F=F_2$, which in combination with
equation~(C8) implies that

$${\cal P}[F|\Gamma_1]={\cal P}[F_1|\Gamma_1]={\cal P}[F_2|\Gamma_2]=
{\cal P}[F|\Gamma_2]\,.\eqno\rm(C10)$$

\@This is the result that we intended to prove.

\bigskip
\@{\bf Appendix D: The variance $\langle F^2({\bf x}) | \Gamma
\rangle$ of the residual field $F({\bf x})$.}
\smallskip
\@A derivation will be given for the expression for the variance $\langle
F^2({\bf x}) | \Gamma \rangle$ of the residual field belonging to the
constraint set $\Gamma$. The residual field $F({\bf x})$ is the
difference between a field $f({\bf x})$ obeying the constraint set $\Gamma$
and the mean ${\bar f}({\bf x})\equiv\langle F({\bf x}) | \Gamma \rangle$ of
all these fields,

$${\bar f}({\bf x})= \xi_i({\bf x}) \xi_{ij}^{-1} c_j.
\eqno\rm(D1)$$

\@The crucial observation that ${\cal P}\left[F|\Gamma\right]$, the
probability of having a residual field $F({\bf x})$
satisfying a particular set of constraints $\Gamma$, is independent of
the numerical value $c_i$ of the constraints $\Gamma$,

$${\cal P}\left[F|\Gamma_1\right]={\cal P}\left[F|
\Gamma_2\right]\,,\eqno\rm(D2)$$

\@implies that

$$\langle F^2({\bf x}) | \Gamma \rangle=\langle F^2({\bf x}) \rangle,
\eqno\rm(D3)$$

\@where $\langle F^2 \rangle$ is the variance in all possible realizations
of the field, and $\langle F^2 | \Gamma\rangle$ the variance for
the ones that obey the constraint set $\Gamma$. From equation~(D3) we find

$$\eqalign{\langle F^2({\bf x})\rangle
=\int {\cal P}\left[\Gamma\right]\,\langle F^2({\bf x}) | \Gamma \rangle
&=\int {\cal P}\left[\Gamma\right]\,\langle (f({\bf x})-{\bar f}
({\bf x}))^2|\Gamma \rangle\cr
&=\int {\cal P}\left[\Gamma\right]\,\biggl\{\langle f^2({\bf x}) | \Gamma
\rangle - \langle f({\bf x}) | \Gamma \rangle^2 \biggr\},\cr}\eqno\rm(D4)$$

\@where ${\cal P}\left[\Gamma\right]$ is the integrated probability of
all realizations that obey the constraint set $\Gamma$. Evaluation of the
first part of the integral in $\rm(D4)$ yields

$$\eqalign{\int{\cal P}\left[\Gamma\right]\,\langle f^2({\bf x})|\Gamma
\rangle&= \int{\cal P}\left[\Gamma\right]\int{\cal P}\left[f({\bf x})|\Gamma
\right]\,f^2({\bf x})
=\int{\cal P}\left[\Gamma\right]\,{\cal P}\left[f({\bf x})|\Gamma\right]\,
f^2({\bf x})\cr
&= \int{\cal P}\left[f({\bf x})\right]\,f^2({\bf x})
= \langle f^2({\bf x}) \rangle=\sigma_0^2,\cr}\eqno\rm(D5)$$

\@where $\sigma_0^2$ is the general variance of the density field
fluctuations. In the derivation of (D5) we have used the fact that
${\cal P}[f|\Gamma]$ is the product of the probability
${\cal P}\left[\Gamma\right]$ with the conditional probability of having
the field $f({\bf x})$ under the condition that it obeys $\Gamma$,
${\cal P}\left[f|\Gamma\right]$ (equation~13, section 2.2).

To evaluate the second part of the integral we use the expression for
the mean field $\langle f|\Gamma\rangle$ in equation~(D1),

$$\eqalign{\int{\cal P}\left[\Gamma\right]\,\langle f({\bf x}) | \Gamma
\rangle^2 &=\int{\cal P}\left[\Gamma\right]\,\xi_i({\bf x})\xi_{ij}^{-1}c_j
\,c_l\, \xi_{kl}^{-1}\xi_k({\bf x})\cr
&=\xi_i({\bf x})\xi_{ij}^{-1}\,\biggl\{\int\,{\cal P}\left(\Gamma\right)
C_j({\bf x}_j) C_l({\bf x}_l)\biggr\}\,\xi_{lk}^{-1}\xi_k({\bf x}) \cr
&=\xi_i({\bf x})\xi_{ij}^{-1}\,\langle C_j C_l \rangle\,\xi_{lk}^{-1}
\xi_k({\bf x})
\ =\ \xi_i({\bf x})\xi_{ij}^{-1}\xi_{jl}\,\xi_{lk}^{-1}\xi_k({\bf x})
\ =\ \xi_i({\bf x})\xi_{ik}^{-1}\xi_k({\bf x}).\cr}\eqno\rm(D6)$$

\@By inserting equations~(D5) and (D6) into equation~(D4) and using
equation~(D3) we find

$$\langle F^2({\bf x}) | \Gamma \rangle = \sigma_0^2 - \xi_i({\bf x})
\xi_{ij}^{-1}\xi_j({\bf x}),\eqno\rm(D7)$$

\@which is the intended expression.

\bigskip
\@{\bf Appendix E: Shape and orientation of a peak in a random
field.}
\smallskip
\@The second order Taylor expansion of a density field around a peak or dip
at position ${\bf x}_d$ in a density field $f({\bf x})$ is given by
equation~(51), which we repeat here for convenience,

$$f_G({\bf x})=f_G({\bf x}_d)+{1 \over 2}\,\sum_{i,j=1}^3\,
{\partial^2 f_G \over \partial x_i \partial x_j}({\bf x}_d)\,
(x_i-x_{d,i})(x_j-x_{d,j})\,.\eqno\rm(E1)$$

\@This quadratic equation can be written in its canonical form by transforming
to the coordinate system ${\bf x}'=\{x_1',x_2',x_3'\}$ whose axes are aligned
along the eigenvectors of the matrix $\nabla_i \nabla_j f_G$. If the
eigenvalues of $-\nabla_i \nabla_j f_g$ are $\lambda_1$, $\lambda_2$ and
$\lambda_3$, equation~(E1) becomes

$$f_G({\bf x}')=f_G(0)-{1 \over 2}\,\sum_{i=1}^3\,\lambda_i x_i'^2,
\eqno\rm(E2)$$

\@where we have chosen the origin of ${\bf x}'$ to coincide with the position
of the peak or dip. In the case of a peak the $\lambda_i$ have a negative
value, for a dip they have a positive value. From equation~(E2) we see
that the isodensity surface $f_G=F$ is a triaxial ellipsoid whose principal
axes are oriented along the coordinate axes, with semiaxes given
by

$$a_i=\left \lbrack {2 (\nu_d \sigma_0(R_G)-F) \over \lambda_i} \right \rbrack
^{1/2},\qquad i=1,\ldots,3.\eqno\rm(E3)$$

\@In equation~(E3) the central height $f_G({\bf x}_d)$ of the overdensity is
expressed in units of $\sigma_0(R_G)$, i.e. $f_G({\bf x}_d)=\nu_c
\sigma_0(R_G)$.

 From equation~(E3) and the fact that the shape of a triaxial ellipsoid is
fully specified by its two axis ratios $a_{12} \equiv (a_1/a_2)$ and
$a_{13} \equiv (a_1/a_3)$ we can infer that constraints on the shape of the
overdensity result in constraints on the ratio of $\lambda_i$'s,

$$\left(\lambda_2 \over \lambda_1\right)=a_{12}^2\,,\qquad
\left(\lambda_3 \over \lambda_1\right)=a_{13}^2\,.\eqno\rm(E4)$$

\@The actual magnitude of the $\lambda_i$'s depends on the steepness of the
density profile around the peak. This steepness is specified by the Laplacian
$\nabla^2 f_G$, as can be observed from the expansion of the density profile
equation~(E2) in spherical coordinates $(x,\theta,\varphi)$,

$$f_G({\bf x})=f_G({\bf x}_d)+\nabla^2 f_G({\bf x}_d)\,{x^2 \over 2}\,
\left\{ 1+A(\theta,\varphi) \right\}\,.\eqno\rm(E5)$$

\@$A(\theta,\varphi)$ is a function of the direction $(\theta,\varphi)$
that describes the asphericity of the peak via its dependence on the
parameters $\lambda_1$, $\lambda_2$ and $\lambda_3$. In deriving
equation~(E5) we used the relation between the $\lambda_i$ and
$\nabla^2 f_G$

$$\sum_{i=1}^3 \lambda_i = -\nabla^2 f_G ({\bf x}_d),\eqno\rm(E6)$$

\@which can be obtained by double differentiation of equation~(E2). Usually
$\nabla^2 f_G$ is expressed in units of $\sigma_2(R_G)=\langle \nabla^2
f_G\,\nabla^2 f_G \rangle^{1/2}$, i.e. $\nabla^2 f_G(R_G)=-x_d \sigma_2(R_G)$.
The expression for $\lambda_1$ is obtained by combination of
equations~(E4) and (E6),

$$\lambda_1={x_d \sigma_2(R_G) \over (1+a_{12}^2+a_{13}^2)}\,.\eqno\rm(E7)$$

\@Once the value of $\lambda_1$ has been determined, the values of
$\lambda_2$ and $\lambda_3$ are obtained by multiplication of
$\lambda_1$ by $a_{12}^2$ and $a_{13}^2$ respectively (equation~E4).

\vskip 0.5cm
\@The orientation of the peak with respect to
the general coordinate system is described by the three Euler angles
$\alpha$, $\beta$ and $\psi$ (see Goldstein 1980). Here $\beta$ is the
angle between the smallest axis of the ellipsoid and the $z$-coordinate axis,
$\alpha$ the angle between the line of nodes and the $x$-coordinate axis, and
$\psi$ the angle between the largest axis of the ellipsoid and the line of
nodes. The line of nodes is the intersection of the $xy$-plane and the plane
defined by the largest and second largest axis of the ellipsoid.
The transformation matrix $A_{ij}$ (equation~60, Sect. 4.2) is
obtained from this definition of the Euler angles,

$$A=\pmatrix{\ \ \cos\alpha \cos\psi-\cos\beta \sin\alpha \sin\psi&
\ \ \sin\alpha \cos\psi+\cos\beta\cos\alpha\sin\psi&
\ \ \sin\beta \sin\psi\cr
-\cos\alpha \sin\psi-\cos\beta \sin\alpha \cos\psi&
-\sin\alpha \sin\psi+\cos\beta \cos\alpha \cos\psi&
-\sin\beta \cos\psi\cr
\sin\beta \sin\alpha&-\sin\beta \cos\alpha&\cos\beta\cr}.\eqno\rm(E8)$$

\@This matrix describes the transformation from the coordinate system
${\bf x}'$ defined by the principal axes of the ellipsoid to the general
coordinate system ${\bf x}$,

$$x_i'= \sum_{j=1}^3\,A_{ij} (x_j-x_{d,j}),\qquad i=1,\ldots,3\,,\eqno\rm(E9)$$

\@with ${\bf x}_d$ the position of the centre of the peak. Thus, $x_i'^2$
transforms as

$$x_i'^2=\sum_{j=1}^3\sum_{k=1}^3\,A_{ij} A_{ik} (x_j-x_{d,j})(x_k-x_{d,k})\,.
\eqno\rm(E10)$$

\@By inserting this transformation into the expression for the density
profile (eq. E2) we obtain the following quadratic equation for the density
profile in the general coordinate system ${\bf x}$,

$$f_G({\bf x})=f_G({\bf x}_d)-{1 \over 2} \,\sum_{j,k=1}^3 \left\{
\sum_{i=1}^3 \lambda_i A_{ij} A_{ik}\right\} (x_j-x_{d,j})(x_k-x_{d,k})\,.
\eqno\rm(E11)$$

\@Because equation~(E11) is equivalent to equation~(E1) we obtain
the following relationship between the second derivatives of $f_G$
and the
orientation, shape and steepness of the dip or peak in the field $f_G$,

$${\partial^2 f_G \over \partial x_i \partial x_j} = -\sum_{k=1}^3\,
\lambda_k A_{ki} A_{kj}\,,\qquad i,j=1,2,3\,.\eqno\rm(E12)$$

\@This is the expression that we use in section 4.2.

\bigskip
\@{\bf Appendix F: Peak constraint kernels and values.}
\smallskip
\@Here we present the explicit expressions for the 18 peak constraints and
the corresponding kernels ${\hat H}_l({\bf k})$, defined in equation (37),
to give an overview and summary of the results in this paper. The filter
function ${\hat W}({\bf k})$ is taken to be the one corresponding to a
Gaussian filter function with smoothing length $R_G$,

$${\hat W}({\bf k})=e^{-k^2 R_G^2 /2}\,.\eqno\rm(F1)$$

\@The peak constraints are presented in 5 groups. The first group consists
of the peak height constraint $f_G({\bf x}_d)$, the second one of the
three constraints on the first derivative of the field,
$\nabla f_G ({\bf x_d})$, and the third one of the second derivatives
$\nabla_i \nabla_j f_G ({\bf x_d})$. In addition, the fourth group
contains the constraints on the peculiar velocity
of the peak, ${\bf v}_G({\bf x}_d)$, while the fifth group corresponds to
the constraints on the five
components of the shear, $\sigma_{G,ij}({\bf x}_d)$,

\settabs 2 \columns

\vskip 0.5cm

\+$f_G({\bf x}_d)=\nu \sigma_0(R_G)$
&${\hat H}_1({\bf k})\,=\,e^{-k^2 R_G^2/2}\,e^{i {\bf k}\cdot {\bf x}_d}$\cr

\vskip 0.5cm

\+${\displaystyle \partial f_G \over \displaystyle \partial x_1}({\bf x_d}) =0$
&${\hat H}_2({\bf k})\,=\,ik_1\,e^{-k^2 R_G^2/2}\,
e^{i {\bf k}\cdot {\bf x}_d}$\cr
\+${\displaystyle \partial f_G \over \displaystyle \partial x_2}({\bf x}_d) =0$
&${\hat H}_3({\bf k})\,=\,i k_2\,e^{-k^2 R_G^2/2}\,
e^{i {\bf k}\cdot {\bf x}_d}$\cr
\+${\displaystyle \partial f_G \over \displaystyle \partial x_3}({\bf x}_d)=0$
&${\hat H}_4({\bf k})\,=\,ik_3\,e^{-k^2 R_G^2/2}\,
e^{i {\bf k}\cdot {\bf x}_d}$\cr

\vskip 0.5cm

\+${\displaystyle \partial^2 f_G \over \displaystyle \partial x_1^2}({\bf
x}_d)=
-\sum_{k=1}^3 \lambda_k A_{k1} A_{k1}$
&${\hat H}_5({\bf k})\,=\,-k_1^2\,e^{-k^2 R_G^2/2}\,
e^{i {\bf k}\cdot {\bf x}_d}$\cr
\+${\displaystyle \partial^2 f_G \over \displaystyle \partial x_2^2}({\bf
x}_d)=
-\sum_{k=1}^3 \lambda_k A_{k2} A_{k2}$
&${\hat H}_6({\bf k})\,=\,-k_2^2\,e^{-k^2 R_G^2/2}\,
e^{i {\bf k}\cdot {\bf x}_d}$\cr
\+${\displaystyle \partial^2 f_G \over \displaystyle \partial x_3^2}({\bf
x}_d)=
-\sum_{k=1}^3 \lambda_k A_{k3} A_{k3}$
&${\hat H}_7({\bf k})\,=\,-k_3^2\,e^{-k^2 R_G^2/2}\,
e^{i {\bf k}\cdot {\bf x}_d}$\cr
\+${\displaystyle \partial^2 f_G \over \displaystyle \partial x_1 \partial x_2}
({\bf x}_d)=-\sum_{k=1}^3 \lambda_k A_{k1} A_{k2}$
&${\hat H}_8({\bf k})\,=\,-k_1 k_2\,e^{-k^2 R_G^2/2}\,
e^{i {\bf k}\cdot {\bf x}_d}$\cr
\+${\displaystyle \partial^2 f_G \over \displaystyle \partial x_1 \partial
x_3}({\bf x}_d)=-\sum_{k=1}^3 \lambda_k A_{k1} A_{k3}$
&${\hat H}_9({\bf k})\,=\,-k_1 k_3\,e^{-k^2 R_G^2/2}\,
e^{i {\bf k}\cdot {\bf x}_d}$\cr
\+${\displaystyle \partial^2 f_G \over \displaystyle \partial x_2 \partial x_3}
({\bf x}_d)=-\sum_{k=1}^3 \lambda_k A_{k2} A_{k3}$
&${\hat H}_{10}({\bf k})\,=\,-k_2 k_3\,e^{-k^2 R_G^2/2}\,
e^{i {\bf k}\cdot {\bf x}_d}$\cr

\vskip 0.5cm

\+$g_{G,1}({\bf x}_d)={\tilde g}_{1} \sigma_{g,pk}(R_G)$
&${\hat H}_{11}({\bf k})\,=\,{\displaystyle 3 \over \displaystyle 2} \Omega H^2
\,
{\displaystyle i k_1 \over \displaystyle k^2}\,e^{-k^2 R_G^2/2}\,
e^{i {\bf k}\cdot {\bf x}_d}$\cr
\+$g_{G,2}({\bf x}_d)={\tilde g}_{2} \sigma_{g,pk}(R_G)$
&${\hat H}_{12}({\bf k})\,=\,{\displaystyle 3 \over \displaystyle 2} \Omega H^2
\,
{\displaystyle i k_2 \over \displaystyle k^2}\,e^{-k^2 R_G^2/2}\,
e^{i {\bf k}\cdot {\bf x}_d}$\cr
\+$g_{G,3}({\bf x}_d)={\tilde g}_{3} \sigma_{g,pk}(R_G)$
&${\hat H}_{13}({\bf k})\,=\,{\displaystyle 3 \over \displaystyle 2} \Omega H^2
\,
{\displaystyle i k_3 \over \displaystyle k^2}\,e^{-k^2 R_G^2/2}\,
e^{i {\bf k}\cdot {\bf x}_d}$\cr

\vskip 0.5cm

\+$E_{G,11}({\bf x}_d)={\tilde \epsilon}\,\sigma_E(R_G)\,
\sum_{k=1}^3 {\cal Q}(\varpi) T_{k1} T_{k1}$
&${\hat H}_{14}({\bf k})\,=\,{\displaystyle 3 \over \displaystyle 2} \Omega H^2
\,
\left({\displaystyle k_1^2 \over \displaystyle k^2}-
{\displaystyle 1 \over \displaystyle 3}\right)\,
e^{-k^2 R_G^2/2}\,e^{i {\bf k}\cdot {\bf x}_d}$\cr
\+$E_{G,22}({\bf x}_d)={\tilde \epsilon}\,\sigma_E(R_G)\,
\sum_{k=1}^3 {\cal Q}(\varpi) T_{k2} T_{k2}$
&${\hat H}_{15}({\bf k})\,=\,{\displaystyle 3 \over \displaystyle 2} \Omega H^2
\,
\left({\displaystyle k_2^2 \over \displaystyle k^2}-
{\displaystyle 1 \over \displaystyle 3}\right)\,
e^{-k^2 R_G^2/2}\,e^{i {\bf k}\cdot {\bf x}_d}$\cr
\+$E_{G,12}({\bf x}_d)={\tilde \epsilon}\,\sigma_E(R_G)\,
\sum_{k=1}^3 {\cal Q}(\varpi) T_{k1} T_{k2}$
&${\hat H}_{16}({\bf k})\,=\,{\displaystyle 3 \over \displaystyle 2} \Omega H^2
\,
\left({\displaystyle k_1 k_2\over \displaystyle k^2}\right)\,
e^{-k^2 R_G^2/2}\,e^{i {\bf k}\cdot {\bf x}_d}$\cr
\+$E_{G,13}({\bf x}_d)={\tilde \epsilon}\,\sigma_E(R_G)\,
\sum_{k=1}^3 {\cal Q}(\varpi) T_{k1} T_{k3}$
&${\hat H}_{17}({\bf k})\,=\,{\displaystyle 3 \over \displaystyle 2} \Omega H^2
\,
\left({\displaystyle k_1 k_3\over \displaystyle k^2}\right)\,
e^{-k^2 R_G^2/2}\,e^{i {\bf k}\cdot {\bf x}_d}$\cr
\+$E_{G,23}({\bf x}_d)={\tilde \epsilon}\,\sigma_E(R_G)\,
\sum_{k=1}^3 {\cal Q}(\varpi) T_{k2} T_{k3}$
&${\hat H}_{18}({\bf k})\,=\,{\displaystyle 3 \over \displaystyle 2} \Omega H^2
\,
\left({\displaystyle k_2 k_3 \over \displaystyle k^2}\right)\,
e^{-k^2 R_G^2/2}\,e^{i {\bf k}\cdot {\bf x}_d}$\cr

}
\end